\documentclass[ALICE,manyauthors]{cernphprep}
\pdfoutput=1
\usepackage[comma,square,numbers,sort&compress]{natbib}
\usepackage{hyperref}
\usepackage{lineno}

\usepackage[usenames,dvipsnames]{color}
\usepackage{graphicx}
\usepackage{amsmath,amssymb}
\graphicspath{{img/}}
\newcommand{\pp}{\ensuremath{\mathrm {p\kern-0.05em p}}}
\newcommand{\PbPb}{\ensuremath{\mbox{Pb--Pb}}}
\newcommand{\pPb}{\ensuremath{\mbox{p--Pb}}}

\newcommand{\sqrtS}{\ensuremath{\sqrt{s}}~}
\newcommand{\sqrtSnn}{\ensuremath{\sqrt{s_{\mathrm{NN}}}}~}
\newcommand{\sqrtSE}[2][TeV]{$\sqrtS = #2\,\mathrm{~#1}$}
\newcommand{\sqrtSnnE}[2][TeV]{$\sqrtSnn = #2\,\mathrm{~#1}$}
\newcommand{\pt}{\ensuremath{p_{\mathrm{T}}}}
\newcommand{\meanpt}{\ensuremath{\langle p_{\mathrm{T}}\rangle}}
\newcommand{\dEdx}{\ensuremath{\mathrm{d}E/\mathrm{d}x}}
\newcommand{\dndy}{\ensuremath{\mathrm{d}N/\mathrm{d}y}}

\newcommand{\avgdndeta}{\ensuremath{\langle\mathrm{d}N_{\rm ch}/\mathrm{d}\eta_{\mathrm{lab}}\rangle}}
\newcommand{\avgdndetava}{\ensuremath{\langle\mathrm{d}N_{\rm ch}/\mathrm{d}\eta_{\mathrm{lab}}\rangle_{\vert\eta\vert<0.5}}}
\newcommand{\Nch}{\ensuremath{\langle N_{\mathrm{ch}}\rangle}}
\newcommand{\dndetacube}{\ensuremath{\langle\mathrm{d}N_{\rm ch}/\mathrm{d}\eta_{\mathrm{lab}}\rangle^{1/3}}}
\newcommand{\dndydpt}{\ensuremath{\mathrm{d}^2 N/(\mathrm{d}\pt\mathrm{d}y)}}

\newcommand{\ncoll}{\ensuremath{N_{\mathrm{coll}}}}
\newcommand{\npart}{\ensuremath{N_{\mathrm{part}}}}
\newcommand{\ncollva}{\ensuremath{\langle\mathrm{N_{coll}}\rangle}} 
\newcommand{\y}{\ensuremath{y}}
\newcommand{\Minv}{\ensuremath{M_{\mathrm{inv}}}}
\newcommand{\MinvKpi}{\ensuremath{M_{\mathrm{K\pi}}}}
\newcommand{\MinvKK}{\ensuremath{M_{\mathrm{KK}}}}
\newcommand{\acceff}{\ensuremath{\mathrm{Acc} \times \epsilon}}
\newcommand{\ks}{\ensuremath{{\rm K}^{*0}}} 

\newcommand{\aks}{\ensuremath{\overline{{\rm K}^{*0}}}}
\newcommand{\ph}{\ensuremath{\phi}}
\newcommand{\pion}{\ensuremath{\pi}}
\newcommand{\pip}{\ensuremath{\pi^{+}}}

\newcommand{\pim}{\ensuremath{\pi^{-}}}

\newcommand{\kap}{\ensuremath{{\rm K}^{+}}}
\newcommand{\kam}{\ensuremath{{\rm K}^{-}}}

\newcommand{\pbar}{\ensuremath{\rm\overline{p}}}

\newcommand{\ptophi}{\ensuremath{(\mathrm{p}+\bar{\mathrm{p}})/\phi}}

\newcommand{\ptks}{\ensuremath{p_{\mathrm{T}}(\ks)}}
\newcommand{\ptphi}{\ensuremath{p_{\mathrm{T}}(\ph)}}
\newcommand{\mptks}{\ensuremath{\langle p_{\mathrm{T}}\rangle(\ks)}}
\newcommand{\mptphi}{\ensuremath{\langle p_{\mathrm{T}}\rangle(\ph)}}
\newcommand{\mptp}{\ensuremath{\langle p_{\mathrm{T}}\rangle(p)}}
\newcommand{\MeVc}{\ensuremath{\mathrm{MeV}\kern-0.05em/\kern-0.02em c}}
\newcommand{\MeVcSq}{\ensuremath{\mathrm{MeV}\kern-0.05em/\kern-0.02em c^2}}
\newcommand{\GeVc}{\ensuremath{\mathrm{GeV}\kern-0.05em/\kern-0.02em c}}
\newcommand{\GeVcSq}{\ensuremath{\mathrm{GeV}\kern-0.05em/\kern-0.02em c^2}}

\usepackage{booktabs}

\begin{document}%

\begin{titlepage}
\PHyear{2015}
\PHnumber{326}     
\PHdate{18 December}  
%

\title{Production of K$^{*}$(892)$^{0}$ and $\phi$(1020) in p--Pb collisions at $\sqrt{s_{\mathrm{NN}}}$ = 5.02 TeV}
\ShortTitle{K$^{*}$(892)$^{0}$ and $\phi$(1020) in p--Pb~at $\sqrt{s_{\mathrm{NN}}}$ = 5.02 TeV}   

\Collaboration{ALICE Collaboration\thanks{See Appendix~\ref{app:collab} for the list of collaboration members}}
\ShortAuthor{ALICE Collaboration} 

\begin{abstract}
The production of K$^{*}$(892)$^{0}$ and $\phi$(1020) mesons has been measured in p--Pb collisions at $\sqrt{s_{\mathrm{NN}}}$ = 5.02 TeV.
K$^{*0}$ and $\phi$ are reconstructed via their decay into charged hadrons with the ALICE detector in the rapidity range --0.5 \textless~$y$ \textless~0. The  transverse momentum spectra, measured as a function of the multiplicity, have p$_{\mathrm{T}}$ range from 0 to 15 GeV/$c$ for K$^{*0}$ and from 0.3 to 21 GeV/$c$ for $\phi$. Integrated yields, mean transverse momenta and particle ratios are reported and compared with results in pp collisions at $\sqrt{s}$ = 7 TeV and Pb--Pb collisions at $\sqrt{s_{\mathrm{NN}}}$ = 2.76 TeV. 
In Pb--Pb and p--Pb collisions, K$^{*0}$ and $\phi$ probe the hadronic phase of the system and contribute to the study of particle formation mechanisms by comparison with other identified hadrons. For this purpose, the mean transverse momenta and the differential proton-to-$\phi$~ratio are discussed as a function of the multiplicity of the event. The short-lived K$^{*0}$ is measured to investigate re-scattering effects, believed to be related to the size of the system and to the lifetime of the hadronic phase. 

\end{abstract}
\end{titlepage}
\setcounter{page}{2}

\section{Introduction}
The phase transition predicted by QCD from ordinary matter to a deconfined Quark-Gluon Plasma (QGP) has been studied in high-energy heavy-ion collision (AA) experiments at the Super Proton Synchrotron (SPS) \cite{ref:NA49, ref:NA49-2, ref:NA50, ref:NA50-2, ref:NA57, ref:NA60,  ref:NA60-2,  ref:NA60-3, ref:NA60-4, ref:WA97, ref:WA97-2}, the Relativistic Heavy Ion Collider (RHIC) \cite{ref:phobos, ref:star, ref:phenix, ref:brahms} and the Large Hadron Collider (LHC) \cite{Aamodt:2010pa, Aamodt:2010pb, ref:alice-dNdetaPbPb, Aamodt:2010jd, Aamodt:2011mr, Aad:2010bu, Chatrchyan:2011sx}. 
In this context, hadronic resonances provide an important contribution to the study of particle production mechanisms and the characterisation of the dynamic evolution of the system formed in heavy-ion collisions, during the late hadronic phase.
Results on resonance production in different collision systems at RHIC have been reported in \cite{ref:star-kstar-AApp, ref:star-baryonRsn-AApp, ref:star-rsn-dAu, ref:star-phi, ref:star-kstar, ref:phenix-phi-raa, ref:phenix-phi}. At the LHC, K$^{*}$(892)$^{0}$ and \ph(1020) production have been measured in pp collisions at \sqrtSE{7} by ALICE \cite{ref:alice-rsn-pp7TeV}, ATLAS \cite{ref:atlas-phi} and LHCb \cite{ref:lhcb-phi}, and in pp and \PbPb~collisions at \sqrtSnnE{2.76} by ALICE \cite{ref:alice-rsn-PbPb, ref:alice-phi-fwd}. Results obtained in \pPb~collisions at \sqrtSnnE{5.02} with the ALICE detector are presented in this paper.

Measurements in smaller collision systems such as proton-proton (pp) and proton-nucleus (pA) constitute a reference for the interpretation of the heavy-ion results. 
In addition, proton-nucleus collisions have proven to be interesting in their own right, as several measurements \cite{ref:cms-ridge-pp, ref:cms-ridge-pPb, ref:alice-ridge-pPb, ref:alice-ridge-pPb2, ref:alice-correlations-pPb} indicate that they cannot be explained by an incoherent superposition of \pp~collisions, but suggest instead the presence of collective effects \cite{ref:hydro-pPb, ref:hydro-corr-pPb}. 
In heavy-ion collisions, the presence of a strong collective radial flow reveals itself in the evolution with centrality of the transverse momentum spectra of identified hadrons \cite{ref:alice-id-PbPb}.
The spectral shapes of of \ks~and \ph~follow the common behaviour found for all the other particles and exhibit an increase of the mean transverse momentum, dominated by the low \pt~region of the spectra where particle production is more abundant, with centrality \cite{ref:alice-rsn-PbPb}.
In central \PbPb~events, particles with similar mass such as the \ph~meson and the proton have similar \meanpt~and, in addition, the \ph/p~ratio as a function of \pt~is flat for \pt~\textless~4 \GeVc. Both observations are consistent with expectations from hydrodynamic models, where the mass of the particle drives the particle spectral shapes at low momenta \cite{ref:hydro-PbPb}. 
On the other hand, in most peripheral \PbPb~collisions, as well as in \pp, the \ph/p ratio exhibits a strong \pt~dependence, suggesting that the production of low and intermediate momentum baryons and mesons occurs by means of other mechanisms such as fragmentation or recombination \cite{ref:fragment, ref:fragment-coal}. 

Similarly to \PbPb, one is interested in searching for collective effects in \pPb~collisions and in studying particle production as a function of the hadron multiplicity, which strongly depends on the geometry of the collision. In this respect, \pPb~collisions provide us with a system whose size in terms of average charged-particle density and number of participating nucleons is intermediate between pp and peripheral \PbPb~collisions \cite{ref:alice-dNdeta-pp, ref:alice-id-pPb, ref:alice-dNdetaPbPb, ref:alice-dndeta-pPb, ref:alice-centrality-pPb}. Measurements in an intermediate-size system as \pPb~can provide information on the onset of the collective behaviour leading to the presence of radial flow. 
\\The \ph~meson, with similar mass as that of the proton and rather long lifetime ($\tau_{\ph}$ = 46.3 $\pm$ 0.4 fm/$c$ \cite{ref:pdg}) compared to that of the fireball, is an ideal candidate for such study. 
The yields of short-lived resonances such as the \ks~($\tau_{\ks}$ = 4.16 $\pm$ 0.05 fm/$c$ \cite{ref:pdg}) instead, may be influenced by interactions during the hadronic phase: the re-scattering of the decay products in the fireball may prevent the detection of a fraction of the resonances, whereas pseudo-elastic hadron scattering can regenerate them. 
The effects of re-scattering and regeneration depend on the scattering cross section, the particle density, the particle lifetime and the timespan between chemical and kinetic freeze-out, namely the lifetime of the hadronic phase. Therefore, the observation of re-scattering effects would imply the presence of an extended hadronic phase. 
The latter can be studied by comparing particles with different lifetimes, such as the \ks~resonance and the \ph~meson, which has a ten times longer lifetime. 
ALICE has observed \cite{ref:alice-rsn-PbPb} that in most central \PbPb~collisions at the LHC the \ks/K ratio is significantly suppressed with respect to peripheral \PbPb~collisions, pp collisions and the value predicted by a statistical hadronisation model \cite{ref:thermal}. This is interpreted as a scenario where re-scattering during the hadronic phase, dominating for low-momentum resonances (\pt~\textless~2 \GeVc) \cite{ref:urqmd, ref:epos3}, reduces the measurable yield of \ks. No suppression is observed instead for the ten times longer-lived \ph, since it decays mainly after kinetic freeze-out. Based on these observations, a lower limit of 2 fm/$c$ on the lifetime of the hadronic phase in 0--20$\%$ most central \PbPb~events could be estimated \cite{ref:alice-rsn-PbPb}.
The \ks~suppression exhibits a monotonic trend with centrality, suggesting a dependence on the volume of the particle source at the kinetic freeze-out. A similar measurement of resonance production as a function of the system size in \pPb~can provide information about the lifetime of the hadronic fireball produced in such a smaller system.

The \ks~and \ph~mesons are reconstructed using the ALICE detector in \pPb~collisions at \sqrtSnnE{5.02}. Their yields, mean transverse momenta and ratios to identified long-lived hadrons in \pPb~collisions are studied as a function of the system size or the multiplicity of the event, and compared with pp and \PbPb. The experimental conditions are briefly presented in Sec.~\ref{sec:alice}. Section \ref{sec:analysis} illustrates the analysis procedure, including event and track selection, signal extraction, efficiency correction and systematic uncertainties. The results are presented in Sec.~\ref{sec:results} and in Sec.~\ref{sec:conclusions} the conclusions are summarised.

\section{Experimental setup}\label{sec:alice}
A complete description of the ALICE detector and its performance during the LHC Run I are reported in \cite{ref:alice-jinst} and \cite{ref:alice-perf}, respectively.
\\The analyses presented in this paper have been carried out on a sample of \pPb~collision events at \sqrtSnnE{5.02} collected in 2013. The LHC configuration was such that the lead beam, with energy of 1.58 TeV per nucleon, was circulating in the counter-clockwise direction, namely towards the ALICE ``A'' side (positive rapidity direction), while the 4 TeV proton beam was circulating in the clockwise direction, towards the ALICE muon spectrometer, or ``C'' side. According to this convention for the sign of the coordinates, the nucleon-nucleon center-of-mass system was moving in the laboratory frame with a rapidity of \y$_{\mathrm{NN}}$ = -0.465 in the direction of the proton beam. 
In the following, \y$_{\mathrm{lab}}$ ($\eta_{\mathrm{lab}}$) are used to indicate the (pseudo) rapidity in the laboratory reference frame, whereas \y~($\eta$) denotes the (pseudo) rapidity in the nucleon-nucleon center-of-mass reference system. 
\\For the results presented in this paper, a low-luminosity data sample has been analysed, consisting of events collected at an hadronic interaction rate of about 10 kHz. The interaction region had a root mean square of 6.3 cm along the beam direction and of about 60 $\mathrm{\mu}$m in the direction transverse to the beam. The event pile-up rate has been estimated to have negligible effects on the results of this analysis. In particular, pile-up of collisions from different bunch crossings is negligible due to the 200 ns bunch-crossing spacing, larger than the integration time of the Zero-Degree Calorimeter (ZDC), while a small fraction of in-bunch pile-up events is removed by the offline analysis, as described in the next section.
\\Small acceptance forward detectors (V0, T0, and ZDC) are used for triggering, event characterisation, and multiplicity studies. 
The trigger is provided by two arrays of 32 scintillator detectors, V0A and V0C, that cover the full azimuthal angle in the pseudo-rapidity regions 2.8 \textless~$\eta_{\mathrm{lab}}$ \textless~5.1 (Pb-going direction) and $-$3.7 \textless~$\eta_{\mathrm{lab}}$ \textless~$-$1.7 (p-going direction), respectively. V0 information is also used to classify events in multiplicity classes (see Sec. \ref{sec:events}). The two quartz Cherenkov detectors T0A (4.6 \textless~$\eta_{\mathrm{lab}}$ \textless~4.9) and T0C ($-$3.3 \textless~$\eta_{\mathrm{lab}}$ \textless~$-$3) deliver the time and the longitudinal position of the interaction. The Zero-Degree Calorimeters (ZDC), consisting of two tungsten-quartz neutron and two brass-quartz proton calorimeters placed symmetrically at a distance of 113 m from the interaction point, on both sides, are used to reject background and to count spectator nucleons.
\\The reconstruction of the primary vertex of the collision and the tracking of particles in the ALICE central barrel is provided by the Inner Tracking System (ITS) and the Time-Projection Chamber (TPC), in the pseudo-rapidity range $\vert\eta_{lab}\vert$~\textless~0.9 and the full azimuthal angle. 
The ITS is a silicon-based detector, constituted by two innermost pixels layers (SPD), two intermediate drift (SDD) and two outer strip layers (SSD), with radii between 3.9 and 43 cm from the beam axis. 
The ALICE main tracker, the TPC, is a 90 m$^{3}$ cylindrical drift chamber filled with Ne-CO$_{2}$ gas and divided in two parts by a central cathode. 
The end plates are equipped with multi-wire proportional chambers whose readout cathode pads allow to sample particle tracks up to 159 points (clusters).
In addition to tracking, the TPC allows particle identification via the specific ionization energy loss \dEdx~in the gas. 
\\The Time-Of-Flight (TOF) detector, a large Multigap Resistive Plate Chamber (MRPC) array covering $\vert\eta\vert$\textless 0.9 and the full azimuthal angle, allows for particle identification at intermediate momenta and has been exploited together with the TPC for the analysis presented in this paper (see Sec. \ref{sec:cuts}).

\subsection{Event selection}\label{sec:events}
The minimum bias trigger during \pPb~data taking was configured to select hadronic events with high efficiency, by requiring a signal in either V0A or V0C.
The resulting sample contains single-diffractive (SD), non-single diffractive (NSD) and electromagnetic (EM) events. Diffractive interactions are described in Regge theory by the exchange of a colour singlet object with the quantum numbers of the vacuum (pomeron). In SD events one of the two nucleons breaks up producing particles in a limited rapidity interval. NSD events include double-diffractive interactions, where both nucleons break up by producing particles separated by a large rapidity gap, and other inelastic interactions. The offline analysis selects events having a coincidence of signals in both V0A and V0C in order to reduce the contamination from SD and EM events to a negligible amount. The trigger and event selection efficiency for NSD events is estimated as $\epsilon_{\mathrm{NSD}}$ = 99.2$\%$ using a combination of Monte-Carlo event generators, as described in \cite{ref:alice-dndeta-pPb, ref:alice-centrality-pPb}. The arrival time of signals on the V0 and the ZDC is required to be compatible with a nominal p-Pb collision occurring close to the nominal interaction point, to ensure the rejection of beam-gas and other machine-induced background collisions.
\\The primary vertex of the collision is determined using tracks reconstructed in the TPC and ITS. In case of low multiplicity events only the information from the SPD is used to reconstruct the vertex, as described in detail in \cite{ref:alice-perf}. 98.5$\%$ of all events have a primary vertex. Minimum bias events with the primary vertex positioned along the beam axis within 10 cm from the center of the ALICE detector are selected offline. A small fraction (0.2$\%$) of pile-up events from the same bunch crossing has been removed from the sample by rejecting events with multiple vertices. Events are accepted if the vertices separately measured by the SPD and using tracks are within 0.5 cm, and if the SPD vertex is determined by at least five track segments defined by one hit in each one of the two layers of the detector.

After the trigger and offline event-selection criteria, the sample used for this analysis counts about 10$^{8}$ events, corresponding to an integrated luminosity of about 50 $\mu$b$^{-1}$. The minimum bias sample has been further divided in several event classes based on the charged-particle multiplicity, estimated using the total charge deposited in the V0A detector positioned along the direction of the Pb beam. \\The yield of \ks~is measured in five multiplicity classes, namely 0--20$\%$, 20--40$\%$, 40--60$\%$, 60--80$\%$ and 80--100$\%$. In case of \ph~seven classes, namely 0--5$\%$, 5--10$\%$, 10--20$\%$, 20--40$\%$, 40--60$\%$, 60--80$\%$ and 80--100$\%$ are used. In addition, minimum bias spectra normalised to the fraction of NSD events are measured for both particles.

In order to study the dependence of particle production on the geometry of the collision, the V0A estimator for the charged particle multiplicity has been used to determine centrality, by following the approach based on the Glauber Monte Carlo model combined with a simple model for particle production \cite{ref:glauber, ref:glaubermc}, a strategy customarily employed in heavy-ion collisions \cite{ref:alice-centrality-PbPb}. The average number of binary collisions \ncollva~(related to the number of participant nucleons \npart~by the simple relation \ncoll~= \npart~-1), obtained with this method for each centrality class, are listed in Tab.~\ref{tab:multiplicity} for future reference, together with the mean charged particle multiplicity density, \avgdndetava~\cite{ref:alice-dndeta-pPb, ref:alice-id-pPb}, here corrected for trigger and vertex-reconstruction inefficiency, which is about 5.5$\%$ in the lowest multiplicity event class. 
In addition, the average \ncoll~has been determined with an hybrid method that uses the ZDC to classify the events according to the energy deposited by the neutrons emitted in the Pb-going direction (by evaporation or fragmentation) or the energy measured with the ZDC in the Pb-going direction
and the assumption that the charged-particle multiplicity measured at mid-rapidity is proportional to the number of participant nucleons. This method was shown \cite{ref:alice-centrality-pPb} to avoid possible bias in the event sample related to the fact that the range of multiplicities used to select a given class in \pPb~collisions is of similar magnitude as the fluctuations on the same quantity. 
The variations of the average \ncoll~for a given multiplicity class, obtained with different methods are found not to exceed 6$\%$ in any of the used classes. 

\begin{table}[htdp]
\begin{center}
\begin{tabular}{ lcc }
\hline
Multiplicity class ($\%$) &  \avgdndetava & \ncollva \\
\hline
    0--5 & 45    $\pm$ 1    & 14.8 $\pm$ 1.5 \\
  5--10 & 36.2 $\pm$ 0.8 & 13.0 $\pm$ 1.3 \\ 
10--20 & 30.5 $\pm$ 0.7 & 11.7 $\pm$ 1.2 \\    
  0--20 & 35.6 $\pm$ 0.8 & 12.8 $\pm$ 1.3 \\
20--40 & 23.2 $\pm$ 0.5 & 9.36 $\pm$ 0.84 \\
40--60 & 16.1 $\pm$ 0.4 & 6.42 $\pm$ 0.46 \\
60--80 &  9.8 $\pm$ 0.2  & 3.81 $\pm$ 0.76\\
80--100& 4.16 $\pm$ 0.09 & 1.94 $\pm$ 0.45 \\ 
\hline
\end{tabular}
\end{center}
\caption{Average charged particle pseudo-rapidity density, \avgdndetava, measured at mid-rapidity in visible cross section event classes  and average number of colliding nucleons, \ncollva. Multiplicity classes are defined using the V0A estimator \cite{ref:alice-dndeta-pPb, ref:alice-centrality-pPb}, as described in the text. Total systematic uncertainties are reported, see \cite{ref:alice-centrality-pPb} for details, that do not include the difference with respect to the other methods used to estimate the average \ncoll. For minimum bias collisions, \avgdndeta~= 16.81 $\pm$ 0.71 and \ncollva~= 6.87 $\pm$ 0.5.}
\label{tab:multiplicity}
\end{table}%

\section{Resonance signal reconstruction}\label{sec:analysis}
K$^{*}$(892)$^{0}$~and \ph(1020) mesons are reconstructed through their decay into charged hadrons, \ks~$\rightarrow$~\kap\pim~and \aks~$\rightarrow$~\kam\pip, B.R. = 0.666, and \ph~$\rightarrow$~\kap\kam, B.R. = 0.489 \cite{ref:pdg}. 
Since K$^{*}$(892)$^{0}$ and $\overline{\mathrm{K}^{*}}$(892)$^{0}$ are expected to be produced in equal amounts, as measured in lower energy experiments \cite{ref:isr}, for this measurement the yields of particle and anti-particle are combined in order to improve statistics. The average (K$^{*}$(892)$^{0}$ + $\overline{\mathrm{K}^{*}}$(892)$^{0}$)/2 is indicated as \ks~in the following. The \ph(1020) meson is indicated as \ph.
\\For these measurements, the reconstructed \ks~and \ph~are selected in the rapidity range $-$0.5~\textless~\y~\textless~0, in order to ensure the best detector acceptance as the center of mass of the nucleon-nucleon system was moving with respect to the beam interaction point.

\subsection{Track selection and particle identification}\label{sec:cuts}
The charged tracks coming from the primary vertex of the collision (``primary'' tracks) with \pt~\textgreater~0.15 \GeVc~and $\vert\eta_{\mathrm{lab}}\vert$ \textless~0.8 are considered for the invariant mass reconstruction of \ks~and \ph~in this analysis. The selection of primary tracks imposes that they satisfy good reconstruction quality criteria. It is required that tracks have left a signal in at least one of the layers of the SPD and that the distance of closest approach to the primary vertex of the collisions is lower than 7$\sigma_{xy}$ in the transverse plane and within 2 cm along the longitudinal direction. The resolution on the distance of closest approach in the transverse plane, $\sigma_{xy}$, is strongly \pt-dependent and lower than 100 $\mu$m for \pt~\textgreater~0.5 \GeVc~\cite{ref:alice-perf}. In addition tracks are required to cross at least 70 out of maximum 159 horizontal segments (or ``rows'') along the transverse readout plane of the TPC.

Primary tracks have been identified as \pion~or K based on the information of the TPC and TOF detectors. In the TPC, charged hadrons are identified by measuring the specific ionisation energy loss (\dEdx) in the detector gas. With a resolution ($\sigma_{\mathrm{TPC}}$) on \dEdx~of 6$\%$, the TPC allows a 2$\sigma_{\mathrm{TPC}}$ separation between \pion~and K up to \pt$\sim$0.8 \GeVc~and above 3 \GeVc, in the relativistic rise region of the \dEdx. The TOF contributes to particle identification with the measurement of the time-of-flight of the particle, with the start time of the event measured by the T0 detector or using an algorithm which combines the particle arrival times at the TOF surface. In p-Pb collisions, when the event time is determined by the TOF algorithm (available for 100 $\%$ of the events which have more than three tracks) the resolution is 80 \textless~$\sigma_{\mathrm{TOF}}$ \textless~100 ps. TOF allows a 2$\sigma_{\mathrm{TOF}}$ separation between identified \pion~and K in the momentum range 0.7--3 \GeVc, and between K and protons up to 5 \GeVc~\cite{ref:tof-perf}. 

\begin{figure}[ht]
\begin{center}   
\includegraphics[width=0.825\textwidth]{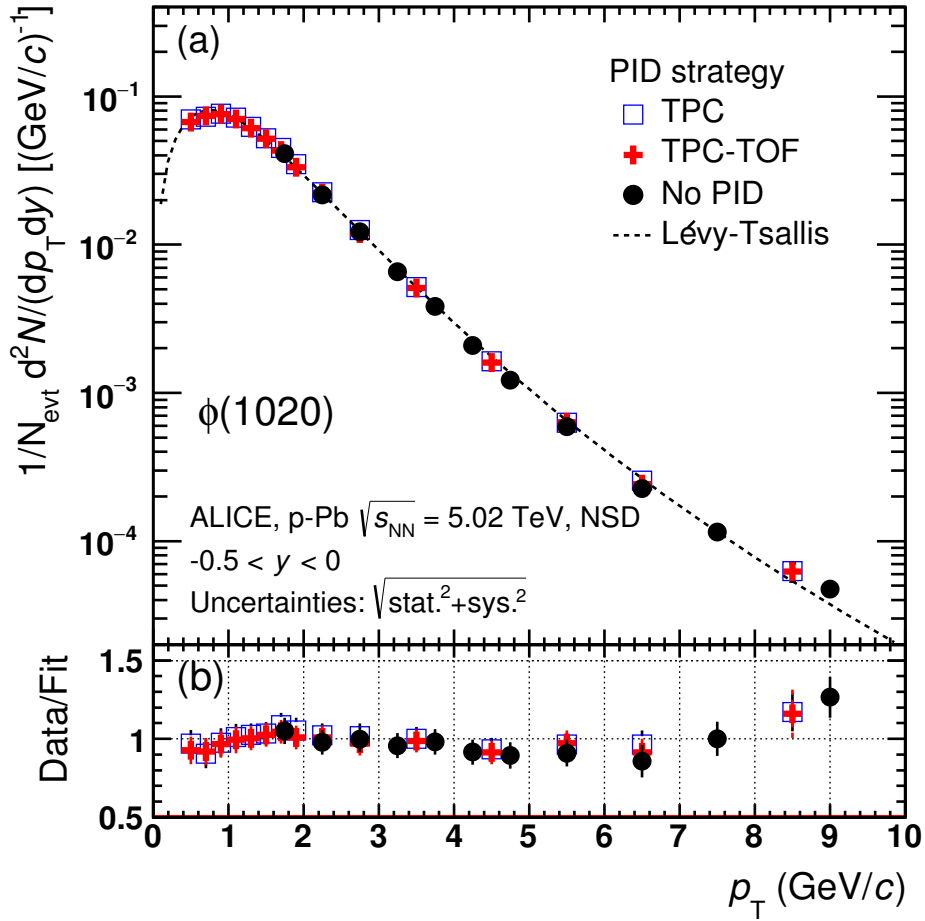} 
\caption{(a) Comparison of the transverse momentum spectrum \dndydpt~of \ph-meson in non-single diffractive (NSD) \pPb~events, reconstructed via the decay channel into \kap\kam~by exploiting three different strategies for K identification: TPC only, TPC-TOF and no-PID. The reader can refer to Sec.~\ref{sec:cuts} for details on the PID selection and to Sec.~\ref{sec:signal} for a description of the signal extraction procedure. The uncertainties are the sum in quadrature of statistical and systematic. A L\`{e}vy-Tsallis function (see Eq. \ref{eq:levy}) is used to fit the TPC-TOF spectrum in 0.3 \textless~\pt~\textless~10 \GeVc. 
(b) Ratio of each spectrum to the fit function, showing good agreement of the three PID strategies within uncertainties.}
\label{fig:pid}
\end{center}
\end{figure}

For the combined ``TPC-TOF PID" approach, particles with a signal in the TOF are identified by requiring that the measured time-of-flight and energy loss do not deviate from the expected values for each given mass hypothesis by more than 2$\sigma_{\mathrm{TOF}}$ and 5$\sigma_{\mathrm{TPC}}$, respectively. For tracks which do not hit the TOF active region, a 2$\sigma_{\mathrm{TPC}}$ selection on the \dEdx~is applied. Variations of these cuts have been used for systematic studies, as described in Sec.~\ref{sec:syst}. 
Besides the TPC-TOF, the measurement of \ph~has been performed following two alternative strategies, one which exploits a 2$\sigma_{TPC}$ separation on the particle energy loss in the TPC for the K identification, and the second for which no PID cuts are applied. In the no-PID scheme all positively charged hadrons are considered as \kap~whereas all negatively charged hadrons are considered as \kam. The no-PID approach extends the measurement of the yields from \pt~= 10 \GeVc, the upper limit reached by the PID analysis, to 16 \GeVc~(multiplicity dependent) or 21 \GeVc~(minimum bias). At low \pt, the TPC-TOF selection leads to a better separation between signal and background with respect to TPC-only and no-PID, therefore it is used until \pt(\ph)$_{\mathrm{cutoff}}$ = 3 \GeVc. At high momentum, K and \pion~cannot be efficiently separated by TPC-TOF, therefore no-PID is used for \pt(\ph) \textgreater~3 \GeVc~to maximise the total reconstruction efficiency. 
The multiplicity-integrated yields of \ph~(see Sec.~\ref{sec:results}) obtained with the no-PID, TPC only, and TPC-TOF approaches are compared in Fig.~\ref{fig:pid}a in the common transverse momentum interval. 
Details on the signal extraction procedure and efficiency correction are given respectively in Sec.~\ref{sec:signal} and \ref{sec:eff}.
The ratio of the data to the L\`evy-Tsallis function (see Sec.~\ref{sec:pt}) used to fit the TPC-TOF spectrum in the 0.3 \textless~\pt~\textless~10 \GeVc~range (Fig.~\ref{fig:pid}b) further shows a good agreement among the three analyses, within uncertainties. 
In the case of \ks, which is a wide resonance, PID is necessary also at high momentum to reduce the background and therefore the TPC-TOF strategy has been applied in the full kinematic range.

\subsection{Signal extraction}\label{sec:signal}
\begin{figure}[t]
\begin{center}   
\includegraphics[width=0.49\textwidth]{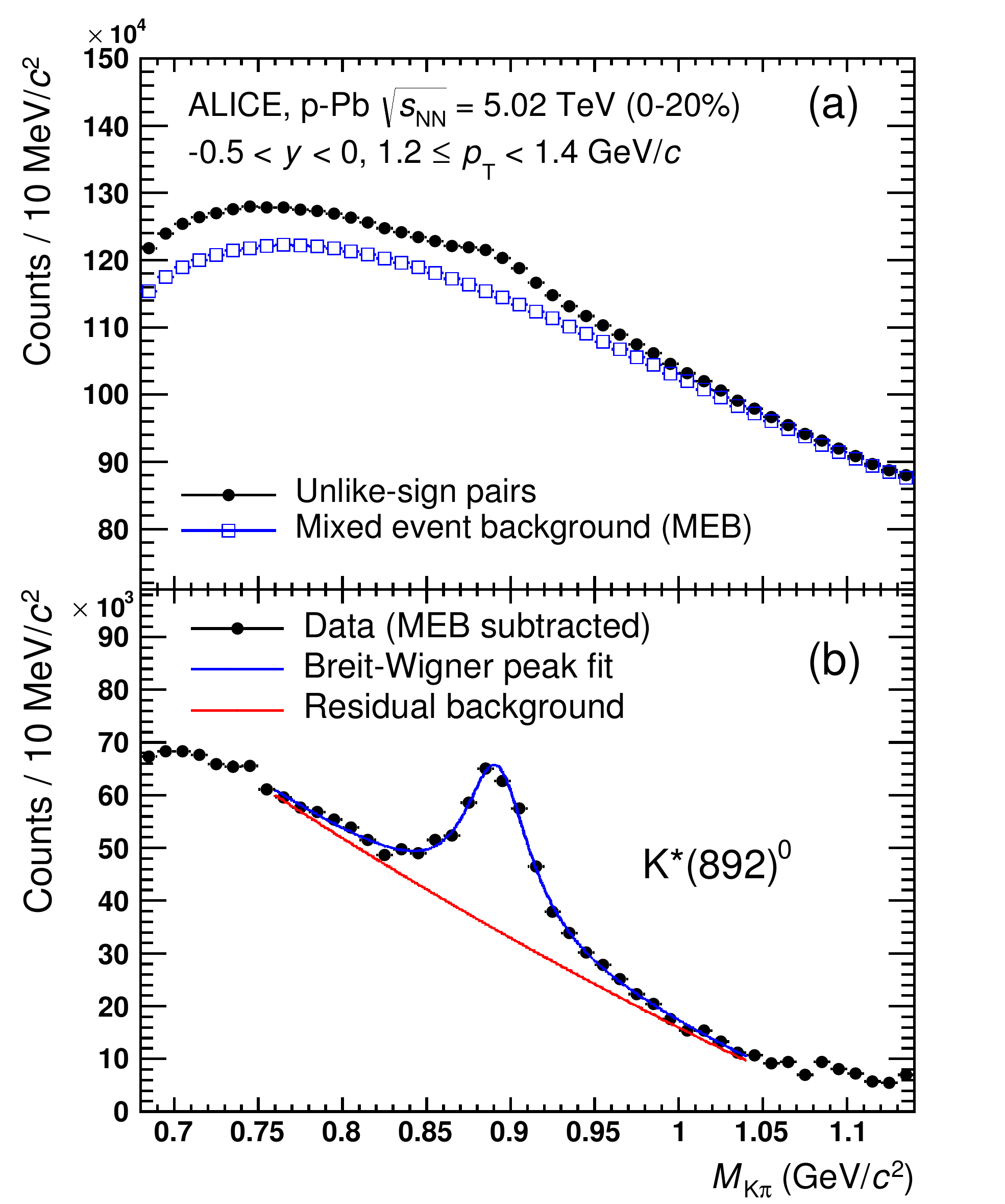}
\includegraphics[width=0.49\textwidth]{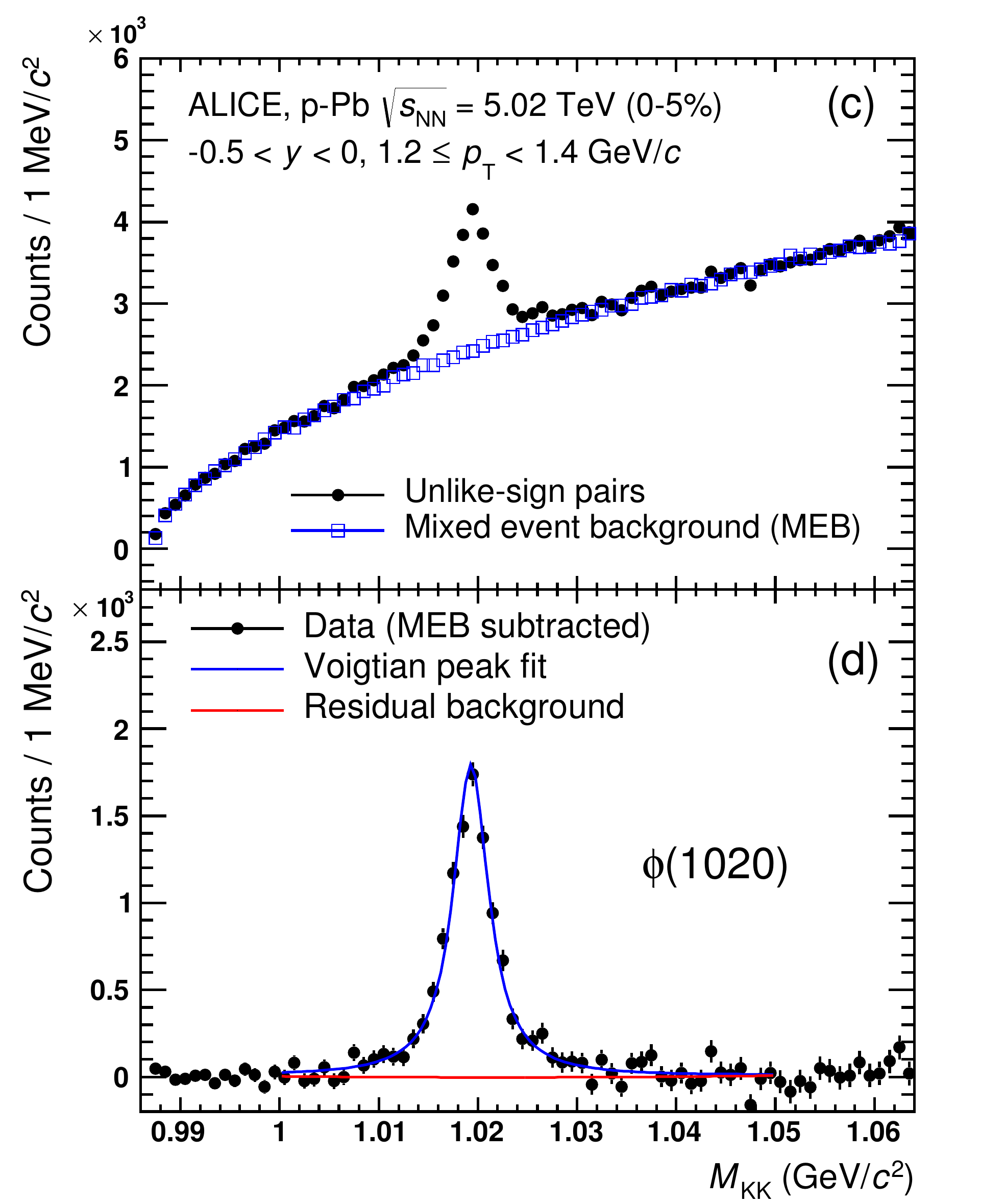}
\caption{Invariant-mass distributions for \ks~(a, b) and \ph~(c, d) in the transverse momentum range 1.2 $\leq$~\pt~\textless~1.4 \GeVc~and multiplicity classes 0--20$\%$ and 0--5$\%$, respectively. Upper panels, (a) and (c), report the unlike-sign invariant-mass distribution and the mixed event background (MEB) normalised as described in the text. In lower panels, (b) and (c), the distributions after background subtraction are shown. The \ks~peak is fitted with a Breit-Wigner function whereas the \ph~meson peak is described by a Voigtian function. A second order polynomial function is used to describe the residual background.}
\label{fig:invmassks}
\end{center}
\end{figure} 

\ks~and \ph~signals are reconstructed in each multiplicity class and transverse momentum interval, as described in \cite{ref:alice-rsn-pp7TeV, ref:alice-rsn-PbPb}. For each event, the invariant-mass distribution of the \ks~(\ph) is constructed using all unlike-sign combinations of charged K candidates with \pion~(K) candidates. For \ks~in the full momentum range and for \ph~up to 3 \GeVc~the TPC-TOF approach has been used for particle identification. \ph~mesons with \pt~\textgreater~3 \GeVc~have been reconstructed by applying no PID. 
In the following the K$^{+}$ and $\pi^{+}$ candidates are labelled as $h^{+}$, the K$^{-}$ and $\pi^{-}$ are labelled as $h^{-}$.
The combinatorial background due to the uncorrelated pairs has been estimated in two ways, by the mixed-event technique and from the invariant-mass distribution of like-sign pairs from the same event. In the event-mixing method the shape of the uncorrelated background is estimated from the invariant-mass distribution of $h^{+}h^{-}$ combinations from five different events. 
Effects from multiplicity fluctuations are minimised by dividing the sample into ten multiplicity classes and by performing event mixing within the same multiplicity class. In order to minimise distortions due to acceptance effects within each multiplicity class, the events are further sub-divided into twenty bins according to the relative vertex position along the $z$-axis ($\Delta{z_{\mathrm{v}}}$ = 1 cm). The final mixed-event distribution for each multiplicity class is found by adding up the \Minv~distributions from each vertex $\Delta{z_{v}}$ interval.  For the \ks~analysis, the mixed-event distribution for each \pt~bin is normalised by the smallest factor that leads to a positive-defined unlike-sign distribution after subtraction, within the statistical error in all invariant-mass bins. The mixed-event distribution for \ph~is normalised in the mass region 1.04 \textless~\MinvKK~\textless~1.06 \GeVcSq. The normalisation range for \ks~and \ph~is varied for systematic studies. 
In the like-sign technique, the invariant-mass distribution for the uncorrelated background is obtained by combining the $h^{+}h^{+}$ and $h^{-}h^{-}$ pairs from the same event according to a geometric mean (2$\sqrt{(h^{+}h^{+})\cdot(h^{-}h^{-})}$), in order to reduce statistical fluctuations in the resulting distribution. The like-sign background is subtracted without normalisation from the unlike-sign pairs distribution.
The mixed-event method has been preferred for \ks~(\ph) signal extraction in the range 0.4 \textless~\pt~\textless~15 \GeVc~(0.3 \textless~\pt~\textless~16 \GeVc), given the smaller statistical uncertainties on the invariant-mass distribution. At very low momentum, \pt~\textless~0.4 \GeVc, the like-sign distribution is found to reproduce better the background of the \ks~and not to be affected by the choice of the normalisation range, therefore it has been preferred over the mixed-event. 
Figure~\ref{fig:invmassks} shows the \MinvKpi~and \MinvKK~invariant-mass distributions before and after background subtraction in the transverse momentum interval 1.2 $\leq$~\pt~\textless~1.4 \GeVc, for the 0--20$\%$ and 0--5$\%$ V0A multiplicity classes, for \ks~and \ph, respectively.  
\\After background subtraction, the resulting distributions exhibit a characteristic peak on top of a residual background (lower panels of Fig.~\ref{fig:invmassks}). The latter is only partly due to imperfections in the description of the combinatorial background and mainly due to correlated pairs from jets, multi-body decay of heavier particles or correlated pairs contribution to the background from real resonance decays where the daughter particles are misidentified as K or \pion~by the TPC-TOF PID. A dedicated study in Monte Carlo simulations has been performed to ensure that the shape of the correlated background is a smooth function of mass and to verify that a second-order polynomial provides a good description of it. 
\\As in \cite{ref:alice-rsn-pp7TeV}, the signal peaks for \ks~and \ph~are fitted respectively with a (non-relativistic) Breit-Wigner and a Voigtian function (convolution of Breit-Wigner and Gaussian) superimposed to a second order polynomial function to shape the residual background. Examples are reported in the lower panels of Fig.~\ref{fig:invmassks}, where fits are performed in the intervals 0.76 \textless~\MinvKpi~\textless~1.04 \GeVcSq~and 1.0 \textless~\MinvKK~ \textless~1.05 \GeVcSq. 
The fitting range is optimised for each \pt~bin across all multiplicity event classes. The mass and width of \ks~and \ph~are found to be compatible with the measurements in \PbPb~collisions \cite{ref:alice-rsn-PbPb}. For the measurement of the yields, the width of \ks~and \ph~have been fixed to their natural values, $\Gamma$(\ks) = 47.4 $\pm$ 0.6 \MeVcSq,  $\Gamma$(\ph) = 4.26 $\pm$ 0.04 \MeVcSq~\cite{ref:pdg}, whereas the resolution parameter of the Voigtian function for \ph~has been kept as a free parameter. The measured \pt-dependent resolution on the \ph~mass (sigma of Gaussian) varies between 0.9 and 1.5 \MeVcSq, and it is consistent with the values extracted from Monte Carlo simulation. The sensitivity to the choice of the normalisation interval, the fitting range, the shape of the background function, the fitting range and the constraints on mass, width and resolution parameters has been studied by varying the default settings, as described in Sec.~\ref{sec:syst}.
\\In minimum bias collisions the sample of reconstructed particles includes about 3.4$\times$10$^{6}$ \ks~and 8.6$\times$10$^{5}$ \ph~in the transverse momentum range 0 \textless~\pt(\ks) \textless~15 \GeVc~and 0.3 \textless~\pt(\ph) \textless~21 \GeVc, respectively. With the available statistics, the \ks~production in the 80--100$\%$ V0A multiplicity event class has been measured up to \pt~= 6 \GeVc, while the \ph~spectra extend up to 16 \GeVc~in the 0-60$\%$ multiplicity percentile interval and up to 13 \GeVc~in 60--80$\%$ and 80--100$\%$.

\subsection{Detector acceptance and efficiency}\label{sec:eff}
\begin{figure}[t]
\begin{center}   
\includegraphics[width=0.7\textwidth]{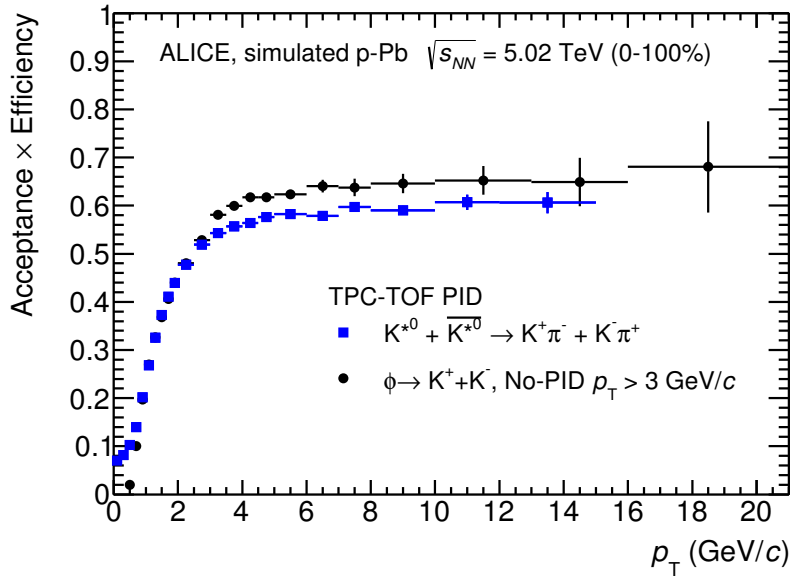}
\caption{Detector acceptance and signal reconstruction efficiency for \ks~and \ph~mesons, which includes reconstruction, track selection and particle identification efficiency. For \ks~and \ph~production below \pt~\textless~3 \GeVc, the PID efficiency is relative to the TPC-TOF approach, whereas for \ph~production with \pt~\textgreater~3 \GeVc~no PID contribution is included, as no particle identification is applied in the analysis.}
\label{fig:efficiency}
\end{center}
\end{figure} 
In order to evaluate the detector acceptance and reconstruction efficiency, a sample of about 10$^{8}$ Monte Carlo simulated \pPb~events, based on the DPMJET 3.05 event generator \cite{ref:dpmjet}, with the detector geometry and material budget modeled by GEANT 3.21 \cite{ref:geant3}, has been analysed. 
The acceptance and efficiency correction is determined as the fraction of generated resonances in the rapidity interval $-$0.5 \textless~\y~\textless~0 that have been reconstructed. The reconstructed signal pairs are obtained upon combination of primary \pion~and K selected by applying the same kinematics cuts and track cuts as in the data (see Sec.~\ref{sec:cuts}), including TPC-TOF PID cuts for \ks, and \ph~with \pt~\textless~3 \GeVc. For \ph~with \pt~\textgreater~3 \GeVc~no PID cuts are applied. The acceptance and efficiency corrections, \acceff, for \ks~and \ph~are reported in Fig.~\ref{fig:efficiency} as a function of \pt~for minimum bias events. 
Since only events with reconstructed primary vertex have been considered in the computation of (\acceff)(\pt), a correction factor has to be applied to the total number of accepted events in each V0A multiplicity event class, to account for vertex reconstruction inefficiency. The correction is about 0.995 for 60-80$\%$ class and 0.945 for the lowest multiplicity events 80-100$\%$, and it is applied as discussed in Sec. \ref{sec:pt}. 

\subsection{Systematic uncertainties}\label{sec:syst}
The measurement of \ks~and \ph~production in \pPb~collisions have been tested for systematic effects due to global tracking efficiency, track selection cuts, PID, signal extraction, knowledge of the material budget and of the hadronic interaction cross section in the detector material, as summarised in Tab.~\ref{tab:sys}. The approach is similar to the one adopted for the study of \ks~and \ph~in \PbPb~collisions \cite{ref:alice-rsn-PbPb}, but the total average uncertainty evaluated in the \pPb~case is significantly lower (about half of the relative uncertainty in the \PbPb), mainly due to lower contributions from global tracking efficiency and the signal extraction procedure. 
No multiplicity dependence of systematic effects has been observed, therefore the uncertainties presented in Sec.~\ref{tab:sys} have been averaged among all multiplicity event classes. For each particle, they are quoted for two separate momentum intervals: for \ks, one can distinguish a low-\pt~range (0 \textless~\ptks~\textless~4 \GeVc) where the knowledge of the material budget and hadronic interaction cross section in the detector material enter in the systematic uncertainty, as opposite to the high-\pt~range (4 \textless~\ptks~\textless~15 \GeVc) where these contributions are negligible (\textless~0.5$\%$). In the \ph~case, two \pt~intervals are considered, according to the particle identification approach used to identify the decay products, namely the ``TPC-TOF'' and ``No PID'' strategies described in Sec.~\ref{sec:cuts}. The \pt~region where the TPC-TOF PID is applied (\pt~\textless~3 \GeVc), coincides also with the range where effects of material budget and hadronic interaction cross section are relevant for the measurement of \ph~production.
\\The main source of uncertainty, common to \ks~and \ph, comes from the determination of the global tracking efficiency. In \pPb~collisions this contribution has been estimated to be a \pt-independent effect of 3$\%$ for charged particles \cite{ref:alice-dndeta-pPb}, which results in a 6$\%$ effect when any two tracks are combined in the invariant-mass analysis of \ks~and \ph. 
\\The track selection was varied to study systematic effects: the analyses are sensitive to variations of the cuts on the number of crossed rows in the TPC and the distance of closest approach to the primary vertex of the collision. Track selection enters in the total uncertainty with a relative contribution of 2.5 $\%$ for \ks~and about 1.9 to 2.2$\%$ for the \ph~case.
\\At high transverse momentum, namely for \ptks~\textgreater~8 \GeVc~and \ptphi~\textgreater~12 \GeVc, the systematic uncertainties are dominated by the raw yield extraction procedure. This contribution is labelled as ``Signal extraction'' in Tab.~\ref{tab:sys} and it includes the background normalisation region, the choice of the fitting range, the residual background shape and variations of the constraints on the fit parameters.
In addition to the default strategy described in Sec.~\ref{sec:signal}, the mixed-event background distributions for \ks~and \ph~have been normalised in different invariant mass regions that surround, but exclude the signal peaks. The sensitivity of the \ks~(\ph) yield extraction to the fit range has been studied by varying each interval boundary within $\pm$50 \MeVcSq ($\pm$5 \MeVcSq). As alternative to the second order polynomial, a third and first order polynomial functions have been used to fit the residual background.
The measurements for both \ks~and \ph~turned out to be independent on the mass parameters, but not on the constraints on the \ks~width and \ph~mass resolution. Therefore, the \ks~width has been varied by $\pm$50$\%$ for systematic studies, while the \ph~resolution has been varied within the range of values observed in the simulation.
Due to the lower particle multiplicity and the improved PID strategy that has led to a lower residual background after mixed-event background subtraction, the contribution of signal extraction for \ks~is reduced by half in \pPb~with respect to the \PbPb~case, where the uncertainty associated to the choice of the fitting range was larger than 9$\%$ \cite{ref:alice-rsn-PbPb}. 
\\In order to study the effect of the PID selection on signal extraction, the cuts on TOF and TPC have been varied to 3$\sigma$ and 4$\sigma$ with respect to the default settings described in Sec.~\ref{sec:cuts}, resulting in the average contribution to the systematic uncertainty reported in Tab.~\ref{tab:sys} as ``particle identification''. 
For \ks~the average contribution from PID is 1.1$\%$ in the low-\pt~range, and 2.7$\%$ at high transverse momenta. The contribution to the \ph~uncertainty is 0.9$\%$ on average in the transverse momentum range where TPC-TOF PID is applied.
\\The knowledge of the material budget contributes for \ks~(\ph) with an average of 1.2$\%$ (2.2$\%$) at low transverse momentum, and a maximum of 3.5$\%$ (5.4$\%$), reached for 0 \textless~\pt~\textless 0.2 \GeVc~(0.8 \textless~\pt~\textless 0.9 \GeVc). In both cases, it is negligible for \pt~\textgreater~3 \GeVc. The contribution from the estimate of the hadronic interaction cross section in the detector material is 1.9$\%$ (2.4$\%$) for \ks~(\ph) at low \pt, negligible for \pt~\textgreater~4 \GeVc~(\pt~\textgreater~3 \GeVc). These effects were evaluated by combining the uncertainties for a $\pion$ and a K (for \ks), and for two K (in case of \ph), determined as in \cite{ref:alice-id-pPb, ref:alice-id-PbPb}, according to the kinematics of the decay. 
\\The systematics were studied independently for all event classes, in order to separate the sources which are multiplicity-dependent and uncorrelated across multiplicity bins. In particular, signal extraction and PID are fully uncorrelated sources, whereas global tracking, track cuts, material budget and hadronic cross section are correlated among different event classes.

\begin{table}[ht]
\small
\begin{center}
\setlength{\tabcolsep}{16pt}
\begin{tabular}{@{} l  c c c c @{}}
\hline
                                                        & \multicolumn{2}{c}{\ks}         & \multicolumn{2}{c}{\ph} \\
\hline  
\pt~(\GeVc)                                    & 0 -- 4.0      & 4.0 -- 15.0           & 0.3 -- 3.0       & 3.0 -- 21.0 \\
PID technique                                & \multicolumn{2}{c}{TPC-TOF}& TPC-TOF   & No PID       \\
\hline  
Global tracking efficiency               & \multicolumn{2}{c}{6$\%$}     & \multicolumn{2}{c}{6$\%$}\\
Track selection cuts                       & \multicolumn{2}{c}{2.5$\%$}   &  1.9$\%$ & 2.2$\%$  \\ 
Material budget                              &  1.2$\%$ & \textless 0.5$\%$ &  2.2$\%$ & \textless 0.5$\%$ \\
Hadronic interaction cross section &  1.9$\%$ & \textless 0.5$\%$ &  2.4$\%$ & \textless 1$\%$ \\
Particle identification                      &  1.1$\%$ & 2.7$\%$                &  0.9$\%$ &  --  \\
Signal extraction                            &  3.8$\%$  & 4.6$\%$               & 1.8$\%$  & 4.3$\%$   \\
\hline
Total                                               &  7.9$\%$  & 8.4$\%$               & 7.4$\%$ & 7.7$\%$ \\
\hline
\end{tabular}
\end{center}
\caption{Sources of systematic uncertainties for \ks~and \ph~yields (\dndydpt). For each source and transverse momentum range (see text for details), the average relative uncertainty over all multiplicity classes is listed. For each \pt~range, the particle identification (``PID technique'') used for the analysis is also indicated. The contributions have been summed in quadrature to estimate the total relative systematic uncertainty.} 
\label{tab:sys}
\end{table}%

\section{Results and discussion}\label{sec:results}

\subsection{Transverse momentum spectra}\label{sec:pt}

\begin{figure}[ht!]
\begin{center}   
\includegraphics[width=0.74\textwidth]{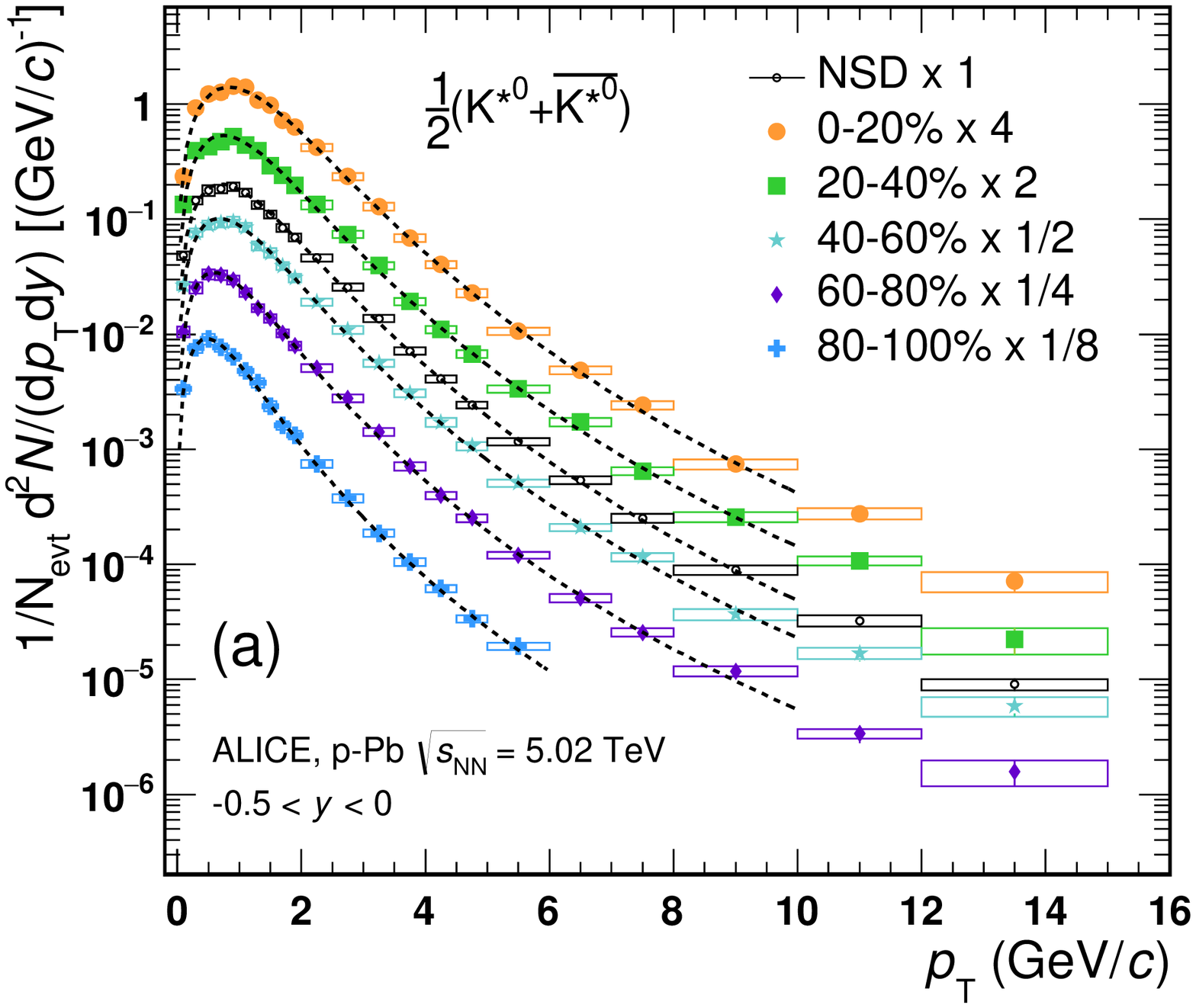} 
\includegraphics[width=0.74\textwidth]{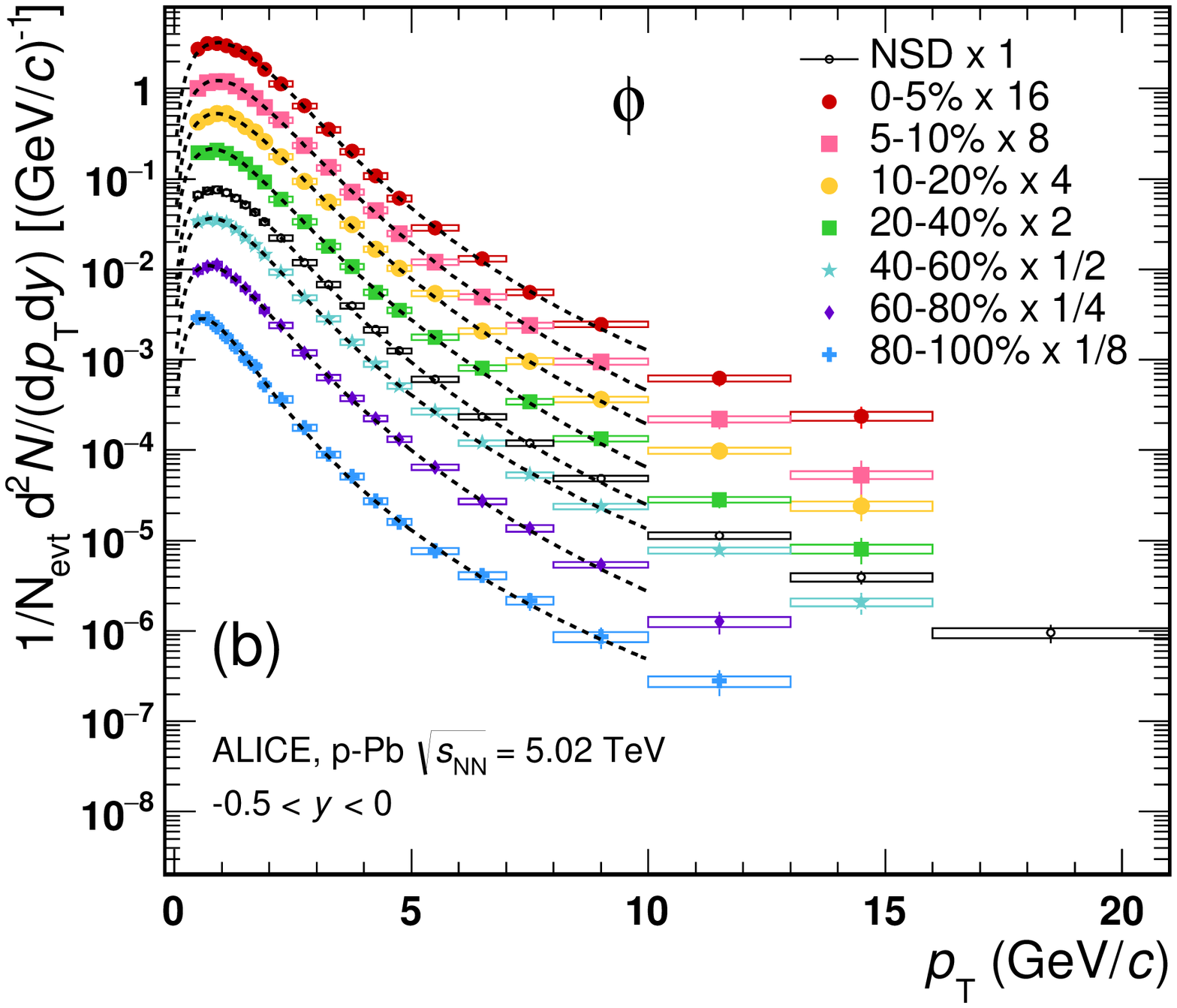} 
\caption{Transverse momentum spectra \dndydpt~of \ks~(a) and \ph~(b) for different multiplicity classes (V0A estimator), measured in the rapidity range $-$0.5 \textless~\y~\textless~0. \ks~and \aks~are averaged. The multiplicity-dependent spectra are normalised to the visible cross section, whereas the minimum bias spectrum is normalised to the fraction of NSD events (see text). Statistical (bars) and systematic (boxes) uncertainties are indicated. Dashed lines represent L\'evy-Tsallis fits, see text for details.}
\label{fig:spectraKstar}
\end{center}
\end{figure}

The multiplicity-dependent transverse momentum spectra of \ks~and \ph~mesons measured in the rapidity range $-$0.5 \textless~$y$ \textless~0 are reported in Fig.~\ref{fig:spectraKstar}. 
Measured yields are corrected for acceptance, efficiency and branching ratio, and normalised to the visible cross section in each V0A multiplicity event class, as discussed in Sec.~\ref{sec:eff}. The minimum bias spectra for \ks~and \ph~are also reported in Fig.~\ref{fig:spectraKstar} and have been normalised to the number of NSD events after applying the correction for trigger efficiency and event selection ($\epsilon_{NSD}$), vertex reconstruction ($\epsilon_{vtx}$) and vertex selection described in Sec.~\ref{sec:alice}, resulting in a total scaling factor of 0.964.

The \pt-integrated particle yields, \dndy, and mean transverse momentum, \meanpt, are determined by using the transverse momentum spectra in the measured range and by using a fit function to extrapolate the yield in the \pt~range where no measurement is available. The same procedure is applied to the spectra of \ks~and \ph~for each event class. 
The L\'evy-Tsallis parameterization \cite{ref:tsallis} has been chosen to fit the corrected \dndydpt~spectra, as it has successfully been adopted to fit the particle spectra in pp collisions at RHIC and at LHC \cite{ref:alice-rsn-pp7TeV, ref:alice-id-pp, ref:STAR-pp-levy, ref:alice-baryonic-rsn-pp}. The L\'evy-Tsallis functional form describes the shape of the exponential spectra at low transverse momentum and the power law distributions at large \pt~with an inverse slope parameter $C$ and an exponent parameter $n$
\begin{equation}
\frac{\mathrm{d}^{2}N}{\mathrm{d}p_{T}\mathrm{d}y} = p_{T}\frac{\mathrm{d}N}{\mathrm{d}y}\frac{(n-1)(n-2)}{nC[nC+m_{0}(n-2)]}\left[1+\frac{\sqrt{p_{T}^{2}+m_{0}^{2}}-m_{0}}{nC}\right]^{-n},
\end{equation}\label{eq:levy}
where $m_{0}$ is the mass of the particle, $n$, $C$ and the integrated yields \dndy~are the free parameters. 
The fits are performed in the \pt~range where the L\'evy-Tsallis function provides a satisfactory description of each spectrum, namely in the interval 0--10 \GeVc~for \ks~and 0.3--10.0 \GeVc~for \ph.
The values of the fit parameters $C$ and $n$, as well as the reduced $\chi^{2}$ are reported in Tab.~\ref{tab:levy}, together with the \dndy~and \meanpt~obtained using the data and the fit function in the extrapolation region. 
\begin{table}[ht]
\begin{center}
\setlength{\tabcolsep}{3.5pt}
\footnotesize
\begin{tabular}{@{}lcccccc@{}}
\hline
\multicolumn{7}{l}{\ks} \\
\hline
Multiplicity ($\%$) &   $C$ (GeV)             &          $n$         & $\chi^{2}$/ndf  & Extr.        &     \dndy~(data + extr.)                                      &  \meanpt~(\GeVc)                 \\ 
\hline
0--20                    &  0.440 $\pm$ 0.010  & 11.1 $\pm$ 0.5  & 1.7 & \textless~10$^{-4}$ & 0.616 $\pm$ 0.008  $\pm$ 0.037 $\pm$ 0.037 & 1.379 $\pm$ 0.011 $\pm$ 0.020 \\ 
20--40                  &  0.430 $\pm$ 0.009  & 9.7 $\pm$ 0.4   & 1.7  & \textless~10$^{-4}$ & 0.426 $\pm$ 0.006 $\pm$ 0.026   $\pm$ 0.026 & 1.300 $\pm$  0.010 $\pm$ 0.019 \\ 
40--60                  &  0.359 $\pm$ 0.008  &  8.8 $\pm$ 0.3  & 0.5 & \textless~10$^{-4}$ & 0.302 $\pm$ 0.004 $\pm$ 0.019  $\pm$ 0.018   & 1.211 $\pm$ 0.009 $\pm$ 0.017 \\ 
60--80                  &  0.309 $\pm$ 0.008  &  7.8 $\pm$ 0.3  & 0.6 & \textless~10$^{-4}$ & 0.185 $\pm$ 0.003 $\pm$ 0.013   $\pm$ 0.011   & 1.108 $\pm$ 0.009 $\pm$ 0.021 \\ 
80--100                &  0.224 $\pm$ 0.008  &  6.2 $\pm$ 0.3  & 0.4 &                       0.002 & 0.083 $\pm$ 0.001 $\pm$ 0.005 $\pm$ 0.005 & 0.943 $\pm$ 0.009 $\pm$ 0.016 \\ 
NSD                     &  0.388 $\pm$ 0.003  &  9.4 $\pm$ 0.1  & 1.8  & \textless~10$^{-4}$ & 0.315 $\pm$ 0.002 $\pm$ 0.018   $\pm$ 0.018   & 1.270 $\pm$ 0.005 $\pm$ 0.017 \\ 
\hline
\multicolumn{7}{l}{\ph}\\
\hline
Multiplicity ($\%$) &   $C$ (GeV)        &    $n$    & $\chi^{2}$/ndf  & Extr.        &                \dndy~(data + extr.)                               &  \meanpt~(\GeVc)  \\ 
\hline 
0--5 	        & 0.472 $\pm$ 0.010          & 12.5 $\pm$ 0.9 & 1.5       & 0.094     & 0.377   $\pm$ 0.004   $\pm$ 0.020   $\pm$ 0.023  & 1.437 $\pm$ 0.009 $\pm$ 0.028 \\ 
5--10        & 0.469 $\pm$ 0.010          & 12.0 $\pm$ 0.8 & 1.1       & 0.094     & 0.288   $\pm$ 0.003   $\pm$ 0.014   $\pm$ 0.017  & 1.442 $\pm$ 0.009 $\pm$ 0.025 \\ 
10--20      & 0.453 $\pm$ 0.010          & 11.3 $\pm$ 0.6 & 1.2       & 0.097     & 0.244   $\pm$ 0.002   $\pm$ 0.012   $\pm$ 0.014  & 1.421 $\pm$ 0.008 $\pm$ 0.024 \\ 
20--40      & 0.413 $\pm$ 0.009          &  9.8 $\pm$ 0.4 & 1.1        & 0.105     & 0.185     $\pm$ 0.001   $\pm$ 0.009 $\pm$ 0.011  & 1.357 $\pm$ 0.006 $\pm$ 0.025 \\ 
40--60      & 0.382 $\pm$ 0.009          &  8.8 $\pm$ 0.4 & 0.6        & 0.115     & 0.1229   $\pm$ 0.0008 $\pm$ 0.0064 $\pm$ 0.0073  & 1.310 $\pm$ 0.006 $\pm$ 0.031 \\ 
60--80      & 0.349 $\pm$ 0.009          &  8.3 $\pm$ 0.4 & 0.5        & 0.115     & 0.0695   $\pm$ 0.0006  $\pm$ 0.0037 $\pm$ 0.0041 & 1.242 $\pm$ 0.008 $\pm$ 0.024 \\ 
80--100    & 0.260 $\pm$ 0.010          &  6.7 $\pm$ 0.3 & 0.4        & 0.163     & 0.0297   $\pm$ 0.0004 $\pm$ 0.0023 $\pm$ 0.0018  & 1.055 $\pm$ 0.010 $\pm$ 0.030 \\ 
NSD        & 0.412 $\pm$ 0.014          & 10.0 $\pm$ 0.5 & 0.8        & 0.106     & 0.1344   $\pm$ 0.0005 $\pm$ 0.0069 $\pm$ 0.0081  & 1.355 $\pm$ 0.003 $\pm$ 0.030 \\ 
\hline
\end{tabular}
\end{center}
\caption{Parameters of the L\'evy-Tsallis fit function and values of \ph~and \ks~\dndy~ and \meanpt~for different multiplicity classes. The $C$ and $n$ parameters with their statistical uncertainty, the reduced $\chi^{2}$ of the fit and the fraction of the total yield obtained by extrapolation (``Extr.") are reported. The yields and \meanpt~are obtained considering data in the measured range and using the result of the fit in the extrapolation region, and are listed as (value $\pm$ stat. $\pm$ uncorr. $\pm$ corr.), where the errors are the statistical uncertainty, the uncorrelated and correlated contributions to the systematic uncertainty, respectively. In the \meanpt~case, the contribution to the systematic uncertainty correlated across multiplicity classes is negligible. The minimum bias spectrum has been normalised to the fraction of non-single diffractive events (NSD) and an additional 3.1 $\%$ relative contribution from the normalisation to NSD has to be considered in the systematic uncertainty on \dndy.}
\label{tab:levy}
\end{table}
\\For \ks~the extrapolation, necessary only at high \pt, covers a fraction of the total yield lower than 0.1$\%$. For \ph~the extrapolated yield is dominated by the fraction of signal in the low transverse momentum region, which constitutes about 10.6$\%$ of the total in the minimum bias case. For all multiplicity classes this fraction is reported in Tab.~\ref{tab:levy}. 
It can be noticed that the inverse slope parameter $C$ and the exponent parameter $n$ increase with multiplicity, reflecting the flattening of the spectra from peripheral to most central events.
\\The uncertainty on \dndy~and \meanpt~is dominated by systematics, which include the contribution of the \pt-uncorrelated systematic uncertainty on the measured spectrum (in average about 6.3$\%$ for \ks,  3.6$\%$ for \ph), the \pt-correlated contributions from global tracking efficiency (6$\%$ for \ks~and \ph, only on \dndy), and the extrapolation of the yield. The first contribution has been estimated by repeating the L\'evy-Tsallis fits moving the measured points within their systematic uncertainties, whereas in order to evaluate the latter, a Blast-Wave function \cite{ref:blastwave} has been used alternatively to fit the spectra. 
The relative systematic uncertainty on \dndy~due to the choice of the fit function varies between 1.5$\%$ and 3$\%$ for \ph, going from high to low multiplicity. Such a contribution is negligible in the case of \ks, due to the fact that its production is measured down to zero transverse momentum.

\subsection{Mean transverse momentum}\label{sec:mpt}
In a hydrodynamically evolving system the spectral shapes are driven by the expansion velocity, thus by the mass of the particle and they are expected to follow ``mass ordering''. Viceversa, the observation of mass ordering of particle spectra may be suggestive of the presence of collective (hydrodynamic) behaviour of the system. Although the presence of a strong radial flow is established in \PbPb~collisions \cite{ref:alice-id-PbPb}, the measurements in \pPb~are not conclusive \cite{ref:alice-id-pPb}, as the comparison between data and models for pp collisions that incorporate final-state effects (such as color-reconnection), shows that the latter could mimic the presence of radial flow. The measurements of \ks~and \ph~can further probe the presence of ``mass ordering'', since they have similar mass as the proton. 

\begin{figure}[bth]
\begin{center}
\includegraphics*[width=0.6\textwidth]{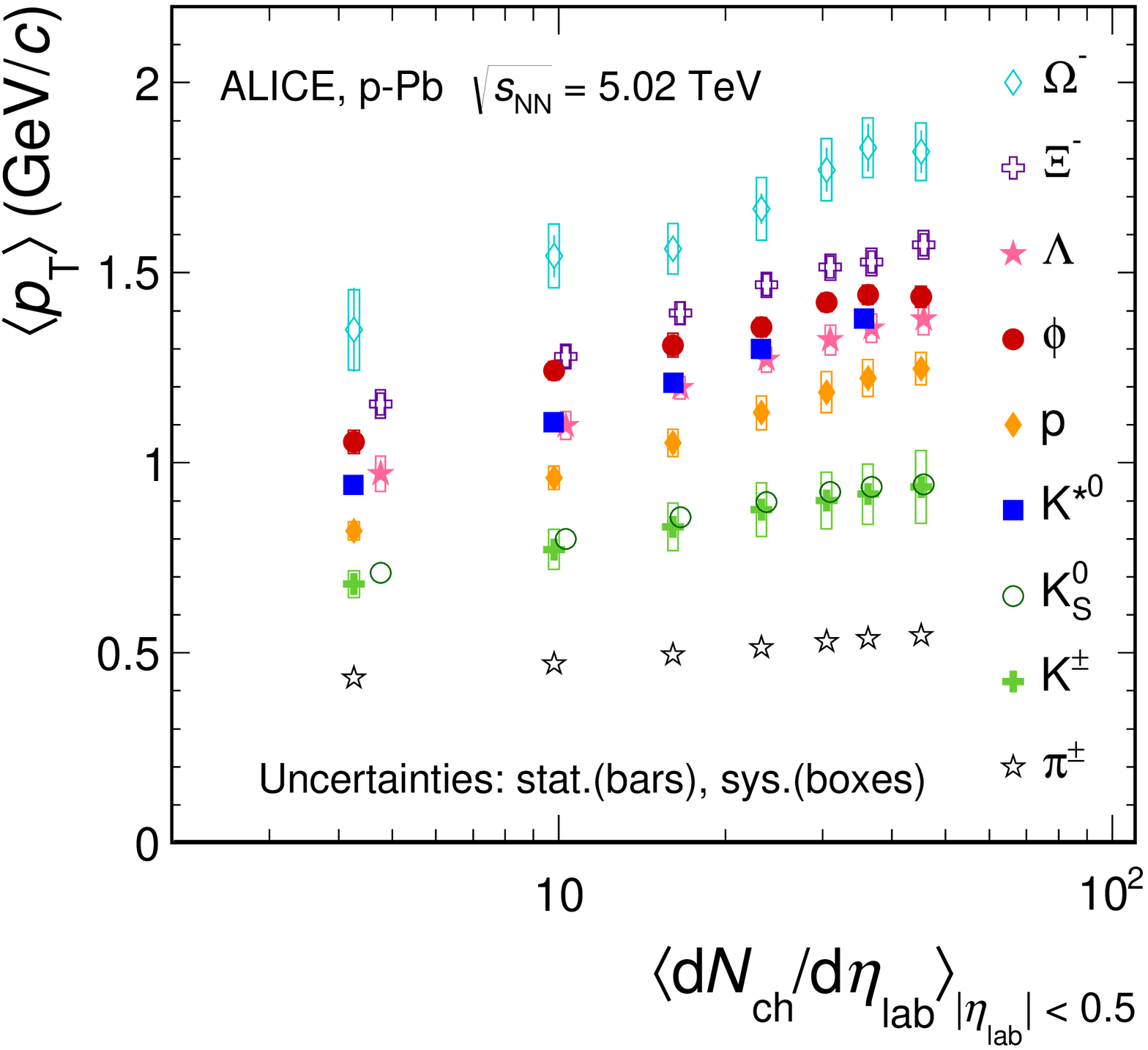}
\caption{Mean transverse momentum of \ks~and \ph~compared to that of identified \pion$^{\pm}$, K$^{\pm}$, K$^{0}_{S}$, p, $\Lambda$, $\Xi^{-}$ and $\Omega^{-}$ previously measured by ALICE in \pPb~collisions at \sqrtSnnE{5.02}~\cite{ref:alice-id-pPb, ref:alice-multis-pPb} as a function of the charged particle density measured in the pseudo-rapidity range $\vert\eta_{\mathrm{lab}}\vert$~\textless~0.5 (\avgdndetava). The K$^{0}_{S}$, $\Lambda$ and $\Xi^{-}$ points are slightly displaced along the x-axis to avoid superposition with other points. Statistical uncertainties are represented as bars, whereas boxes indicate systematic uncertainties.}
\label{fig:meanPt_pPb}
\end{center}
\end{figure}

The transverse momentum spectra of \ks~and \ph, reported in Fig.~\ref{fig:spectraKstar}, become flatter, thus harder, going from the most peripheral to the most central \pPb~events. In other words, the mean transverse momentum increases with multiplicity. 
This is also shown in Fig.~\ref{fig:meanPt_pPb}, where the \meanpt~of \ks~and \ph~as a function of the average charged particle multiplicity density (\avgdndetava) is compared to that of other identified hadrons, including \pion$^{\pm}$, K$^{\pm}$, K$^{0}_{S}$, p, $\Lambda$, $\Xi^{-}$ and $\Omega^{-}$, in \pPb~collisions at \sqrtSnnE{5.02}~\cite{ref:alice-id-pPb, ref:alice-multis-pPb}. 
Going from the lowest to the highest multiplicity events, the relative increase of \meanpt~for \ks~and \ph~mesons is about 40$\%$, common to a wide variety of particles, including K$^{\pm}$, K$^{0}_{S}$, $\Lambda$, $\Xi^{\pm}$ and $\Omega^{\pm}$.  
The relative increase is smaller for \pion~(about 26$\%$) but larger for protons (about 52$\%$). 
The \meanpt~of \ks~is about 10$\%$ larger than that of proton in all event classes and compatible with \meanpt~of $\Lambda$. The \meanpt~of \ph~is instead about 20$\%$ larger than proton and between 4$\%$ (0--5$\%$) and 8$\%$ (80--100$\%$) larger than $\Lambda$. A similar hierarchy is also observed in pp collisions and peripheral \PbPb~collisions, but not in central \PbPb~collisions, where, as expected from hydrodynamics, particles with similar mass have similar \meanpt. 

\begin{figure}[ht!]
\begin{center}
\includegraphics*[width=0.95\textwidth]{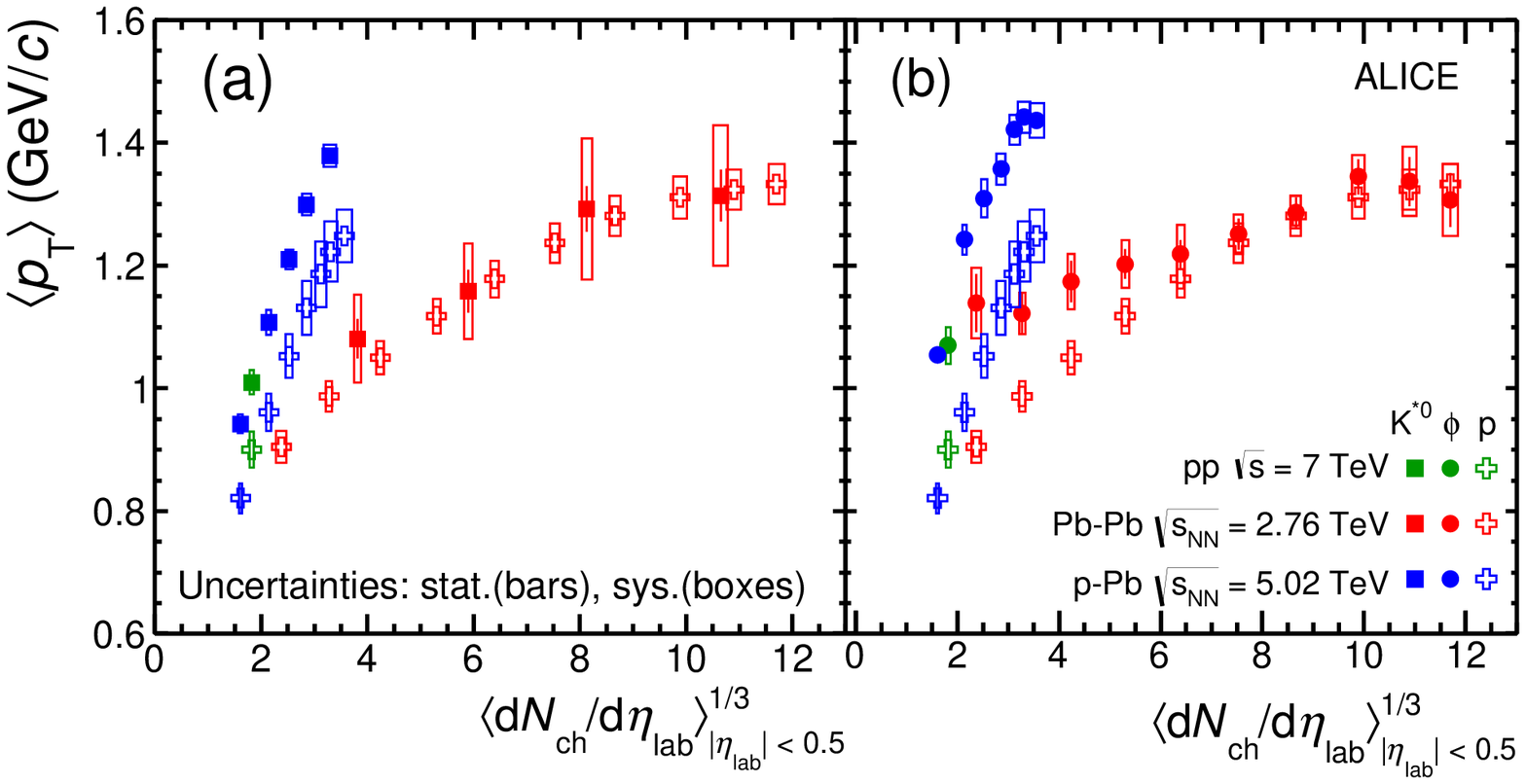}
\caption{System size dependence of the mean transverse momentum of \ks~compared to that of the proton (left panel) and \meanpt~of \ph~compared to that of the proton (right panel). The system size is defined as the cubic root of the average charged particle multiplicity density (\dndetacube) measured in the ALICE central barrel ($\vert\eta_{\mathrm{lab}}\vert$ \textless~0.5) in \pp~at \sqrtS~= 7 TeV (green) \cite{ref:alice-rsn-pp7TeV, ref:alice-id-pp}, \pPb~at \sqrtSnn~= 5.02 TeV (blue) \cite{ref:alice-id-pPb, ref:alice-multis-pPb} and \PbPb~at \sqrtSnnE{2.76}(red) \cite{ref:alice-rsn-PbPb, ref:alice-id-PbPb}. Statistical uncertainties are represented as bars, boxes indicate total systematic uncertainties.}
\label{fig:meanpt}
\end{center}
\end{figure}

\begin{figure}[h!]
\begin{center}
\includegraphics*[width=0.6\textwidth]{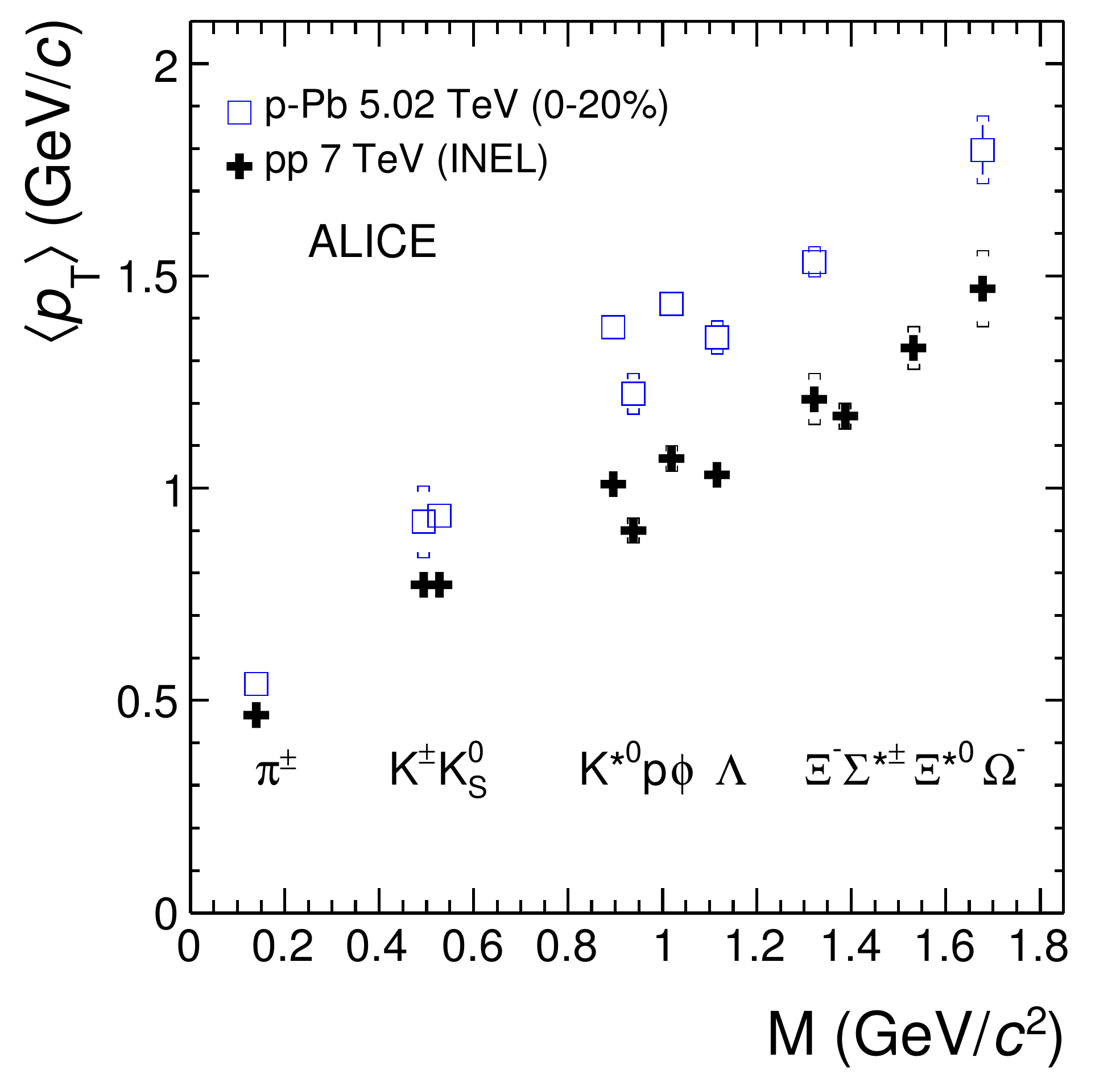}
\caption{Mass dependence of the mean transverse momentum of identified particles measured by ALICE in minimum bias pp collisions at \sqrtSE{7} \cite{ref:alice-id-pp, ref:alice-rsn-pp7TeV, ref:alice-baryonic-rsn-pp} and 0--20$\%$ \pPb~collisions \cite{ref:alice-id-pPb, ref:alice-multis-pPb}. Statistical uncertainties are represented as bars, boxes indicate total systematic uncertainties.}
\label{fig:mptvsmass}
\end{center}
\end{figure}

In Fig.~\ref{fig:meanpt}~the \meanpt~of \ks, proton and \ph~are compared in the three collision systems as a function of the cubic root of the average charged particle multiplicity density, \dndetacube. Based on the observation that the femtoscopic radii scale approximately linearly with \dndetacube~  \cite{Aamodt:2011mr}, this observable is used as a proxy for the system size, associated with the radius of the fireball at  freeze-out. In \pPb, where no extended hadronic medium is expected to be formed, the system size can be associated to the width of the distribution of the particle emission points. The argument holding for \PbPb~has been extended in this paper also to the proton-nucleus case, based on the linear trend of the measured radii with \dndetacube~in \pPb~collisions \cite{Adam:2015pya}. 
From Fig.~\ref{fig:meanpt} one can see that at similar event multiplicity, the \meanpt~is larger in \pPb~than in \PbPb~and the increase with multiplicity is steeper. An analogous observation for unidentified charged particles has been reported in \cite{ref:alicemeanptsystems}, where it is shown that in models of pp collisions, the strong increase in \meanpt~with \Nch~can be understood as the effect of color reconnection between strings produced in multi-parton interactions. 
Considering that for a given multiplicity class in \pPb~and peripheral \PbPb~events the geometry of the collision and the dynamics of the systems are different but the production of \ks~and \ph~mesons relative to long-lived hadrons is comparable (see Sec.~\ref{rec:ratios}), one can conclude that the sample of \pPb~collisions is dominated by events with a larger fraction of quadri-momentum transferred, thus ``harder''.
\\In central \PbPb~collisions, \meanpt~of \ks, proton and \ph~are consistent within uncertainties (Fig.~\ref{fig:meanpt}), and follow ``mass ordering''. This is consistent with the hypothesis that particle boost in the hadronic phase is driven by radial flow \cite{ref:alice-id-PbPb, ref:alice-rsn-PbPb}. This mass ordering seems to weaken going towards peripheral \PbPb~collisions, where it is only approximate. In \pPb~and minimum bias \pp~collisions \mptphi~\textgreater~\mptks~\textgreater~\mptp. 

The \meanpt~for several particles as a function of their mass for 0-20$\%$ \pPb~at \sqrtSnnE{5.02} and minimum bias \pp~collisions at \sqrtSE{7} \cite{ref:alice-id-pp, ref:alice-rsn-pp7TeV, ref:alice-baryonic-rsn-pp} are illustrated in Fig.~\ref{fig:mptvsmass}. In \pPb, the \meanpt~of all particles but \ks~has been obtained as the average between the available measured values weighted by the particle integrated yields in 0-5$\%$, 5-10$\%$ and 10-20$\%$ \cite{ref:alice-id-pPb, ref:alice-multis-pPb}. For \ks, the direct measurement of \meanpt~in 0-20$\%$ is available (see Tab. \ref{tab:levy}). For the \pp~case, also the recent measurements on the short-lived baryonic resonances $\Sigma$(1385)$^{\pm}$ and $\Xi$(1530)$^{0}$ (indicated as $\Sigma^{*\pm}$ and $\Xi^{*0}$) have been included in the comparison. The mean transverse momentum is larger for larger masses, but Fig.~\ref{fig:mptvsmass} shows that in \pp~and \pPb~collisions the \meanpt~values for \ks~and \ph~mesons are systematically larger with respect to a linear trend which includes protons and $\Lambda$ instead. 
These results seem to suggest that a different type of scaling holds in \pp~and \pPb~collisions and prepare the way for a more detailed investigation, which is however outside of the scope of this paper.

\begin{figure}[tb]
\begin{center}
\includegraphics*[width=0.6\textwidth]{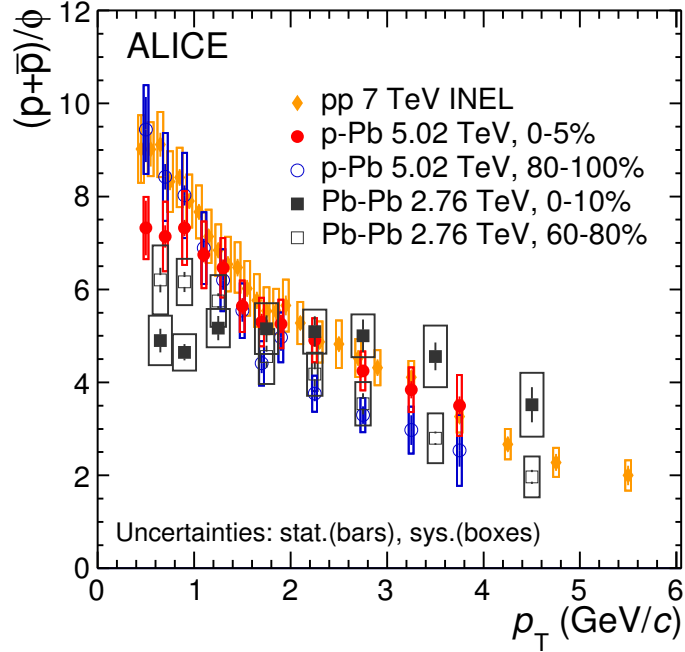}
\caption{The \ptophi~ratio measured in \pPb~in 0--5$\%$ and 80--100$\%$ V0A multiplicity classes, compared to the same ratio measured in minimum bias \pp~collision at \sqrtSE{7} \cite{ref:alice-rsn-pp7TeV, ref:alice-id-pp}, 0--10$\%$ central and 60-80$\%$ peripheral \PbPb~collisions at \sqrtSnnE{2.76} \cite{ref:alice-rsn-PbPb}.}
\label{fig:p2phi}
\end{center}
\end{figure}
\subsection{Differential (p+\textbf{$\overline{\mathrm{p}}$})/$\phi$ ratio}
The multiplicity dependence of the (p+\pbar)/\ph~ratio  as a function of transverse momentum is studied to compare the spectral shapes of \ph~mesons and protons \cite{ref:alice-id-pPb}.
The differential ratios for the 0--5$\%$ and 80--100$\%$ V0A multiplicity event classes in \pPb~collisions are reported in Fig.~\ref{fig:p2phi} together with the ratios in minimum bias pp collisions, 0--10$\%$ central and 60-80$\%$ peripheral \PbPb. 
In peripheral \pPb~the \ptophi~ratio exhibits a qualitatively similar steep decrease as in \pp~collisions, and it is consistent with the ratio measured in 80-90$\%$ peripheral \PbPb~collisions (\cite{ref:alice-rsn-PbPb}, not shown in Fig. \ref{fig:p2phi}). The flat behaviour of \ptophi~for \pt~\textless~3 \GeVc~in 0--10$\%$ central \PbPb~collisions has been previously discussed in \cite{ref:alice-rsn-PbPb} and found to be consistent with the expectations of hydrodynamic models.
In the 0--5$\%$ \pPb, a hint of flattening is observed for \pt~\textless~1.5 \GeVc, but systematic uncertainties are such that no conclusive evidence can be derived. Despite being about 10--20$\%$ larger but compatible within uncertainties, the best qualitative agreement of \ptophi~in high-multiplicity \pPb~(0--5$\%$ V0A multiplicity event class) is achieved with respect to the 60--80$\%$ peripheral \PbPb~collisions, which has also a similar particle multiplicity.  

\subsection{Integrated particle ratios}\label{rec:ratios}
\begin{figure}[ht!]
\begin{center}
\includegraphics*[width=0.63\textwidth]{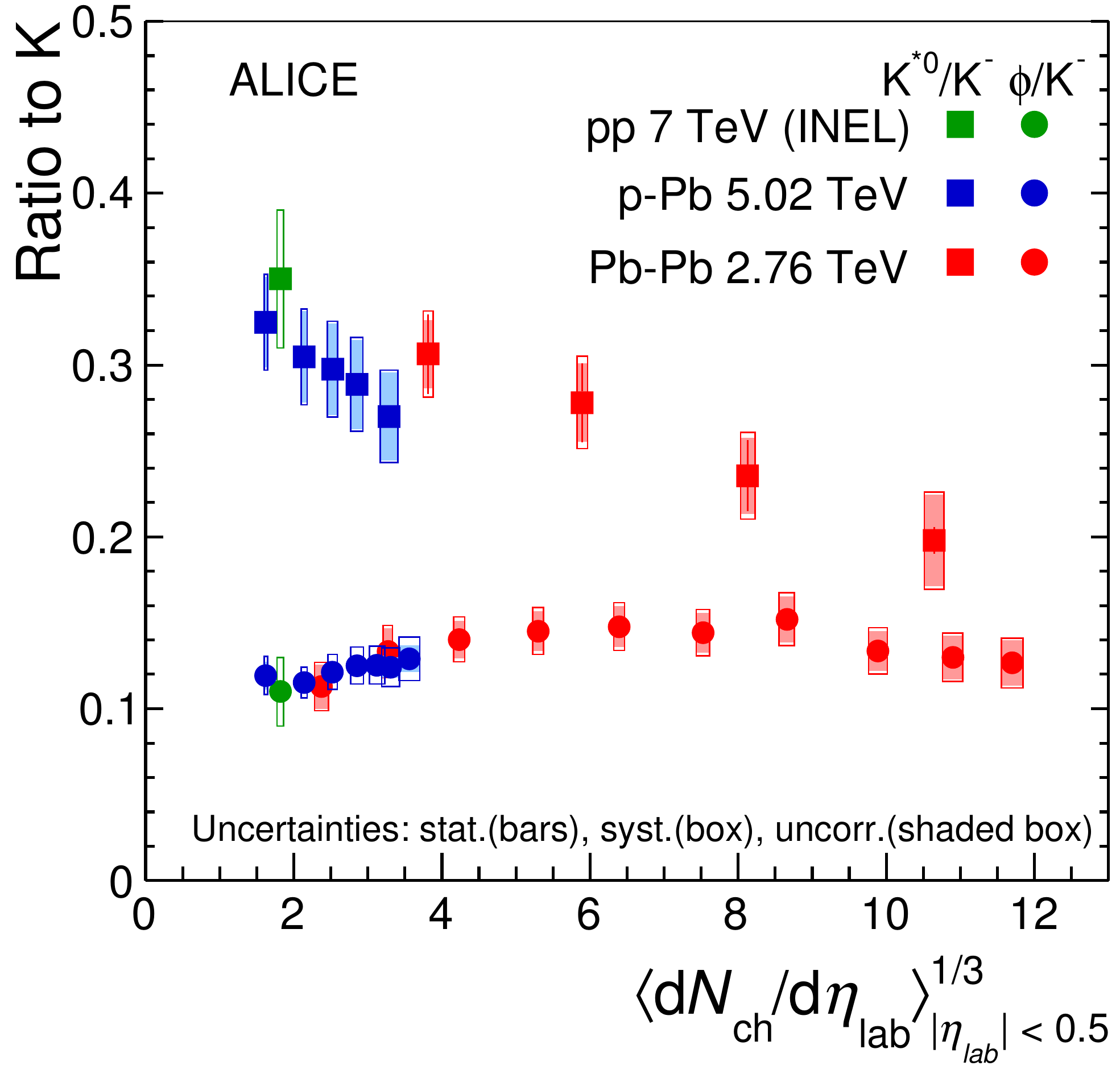}
\caption{Ratio of  \ks~and \ph~to charged K measured in the three collision systems, as a function of the cube root of the average charged particle density (\dndetacube) measured at mid-rapidity, used as a proxy for the system size. Squares represent \ks/K, circles refer to \ph/K. Statistical uncertainties (bars) are shown together with total (hollow boxes) and multiplicity-uncorrelated (shaded boxes) systematic uncertainties. Measurements in pp at \sqrtSE{7} and \PbPb~collisions  at \sqrtSnnE{2.76} are taken from \cite{ref:alice-rsn-pp7TeV} and \cite{ref:alice-rsn-PbPb}, respectively.}
\label{fig:rsn2ka}
\end{center}
\end{figure}

Particle ratios are useful observables to study particle production mechanisms by comparing particles with similar or different strangeness content, mass and lifetime. Short-lived particles such as \ks~and \ph~are used in heavy-ion collisions to derive information on the lifetime of the hadronic phase and on the mechanisms which take place before the kinetic freeze-out, such as re-scattering and regeneration.
If dominant over regeneration, re-scattering is expected to reduce the observed yield of resonances, especially at low momentum and in high particle density environments \cite{ref:urqmd}. For the \ks~resonance re-scattering is the dominant effect at play in most central \PbPb~collisions (and at low transverse momentum, \pt~\textless~2 \GeVc). This observation comes from the strong centrality dependence of the \ks/K ratio (see Fig.~\ref{fig:rsn2ka}) and its direct comparison with the ratio of the longer-lived \ph~meson relative to K \cite{ref:alice-rsn-PbPb}.
\\For \pPb~collisions, the ratios of \ks~and \ph-meson production to that of long-lived hadrons have been computed starting from the integrated yields of \pion, K and proton measured by ALICE with the same data sample \cite{ref:alice-id-pPb}, and are reported for each multiplicity class in Tab.~\ref{tab:ratios}. 
\begin{table}[h!]
\begin{center}
\footnotesize
\setlength{\tabcolsep}{4.5pt}
\begin{tabular}{@{}lccc@{}}
\hline
\multicolumn{4}{l}{\ks}\\
\hline
Multiplicity ($\%$) & (\ks+\aks)/(\pip+\pim)                       &  (\ks+\aks)/(\kap+\kam)             & (\ks+\aks)/(p+\pbar) \\
\hline
0--20     & 0.0379 $\pm$ 0.0006 $\pm$ 0.0028 (0.0026) &  0.270 $\pm$ 0.004 $\pm$ 0.027 (0.026) &  0.676 $\pm$ 0.009 $\pm$ 0.062 (0.059) \\ 
20--40   & 0.0392 $\pm$ 0.0006 $\pm$ 0.0029 (0.0027) &  0.289 $\pm$ 0.004 $\pm$ 0.027 (0.026) &  0.698 $\pm$ 0.009 $\pm$ 0.063 (0.060) \\ 
40--60   & 0.0395 $\pm$ 0.0006 $\pm$ 0.0030 (0.0028) &  0.298 $\pm$ 0.004 $\pm$ 0.028 (0.026) &  0.700 $\pm$ 0.009 $\pm$ 0.064 (0.060) \\ 
60--80   & 0.0393 $\pm$ 0.0006 $\pm$ 0.0032 (0.0029) &  0.308 $\pm$ 0.004 $\pm$ 0.028 (0.026) &  0.696 $\pm$  0.009 $\pm$ 0.065 (0.061) \\ 
80--100 & 0.0399 $\pm$ 0.0006 $\pm$ 0.0030 (0.0028) &  0.325 $\pm$ 0.005 $\pm$ 0.028 (0.026) &  0.745 $\pm$  0.011 $\pm$ 0.067 (0.063) \\
\hline
\multicolumn{4}{l}{\ph}\\
\hline
Multiplicity ($\%$) &  2\ph/(\pip+\pim)                               &  2\ph/(\kap+\kam)                        & 2\ph/(p+\pbar) \\
\hline
0--5       & 0.0185 $\pm$ 0.0002 $\pm$ 0.0014 (0.0009) &  0.1290 $\pm$ 0.0013 $\pm$ 0.0126 (0.0076)  &  0.331 $\pm$ 0.003 $\pm$ 0.030 (0.016) \\ 
5--10     & 0.0174 $\pm$ 0.0002 $\pm$ 0.0013 (0.0006) &  0.1241 $\pm$ 0.0013 $\pm$ 0.0112 (0.0057)  &  0.311 $\pm$ 0.003 $\pm$ 0.028 (0.012) \\ 
10--20   & 0.0174 $\pm$ 0.0001 $\pm$ 0.0012 (0.0006) &  0.1254 $\pm$ 0.0010 $\pm$ 0.0110 (0.0053)  &  0.310 $\pm$ 0.003 $\pm$ 0.027 (0.011) \\ 
20--40   & 0.0170 $\pm$ 0.0001 $\pm$ 0.0012 (0.0006) &  0.1250 $\pm$ 0.0008 $\pm$ 0.0107 (0.0053)  &  0.303 $\pm$ 0.002 $\pm$ 0.026 (0.011) \\ 
40--60   & 0.0161 $\pm$ 0.0001 $\pm$ 0.0012 (0.0006) &  0.1213 $\pm$ 0.0009 $\pm$ 0.0102 (0.0052)  &  0.286 $\pm$ 0.002 $\pm$ 0.025 (0.011) \\ 
60--80   & 0.0147 $\pm$ 0.0001 $\pm$ 0.0011 (0.0006) &  0.1143 $\pm$ 0.0012 $\pm$ 0.0089 (0.0046)   &  0.261 $\pm$ 0.003 $\pm$ 0.022 (0.010) \\ 
80--100 & 0.0143 $\pm$ 0.0002 $\pm$ 0.0013 (0.0009) &  0.1160 $\pm$ 0.0018 $\pm$ 0.0110 (0.0078)   &  0.267 $\pm$ 0.004 $\pm$ 0.028 (0.018) \\ 
\hline
\end{tabular}
\end{center}
\caption{Ratio of \ks~resonance and \ph-meson yields to long-lived hadrons \cite{ref:alice-id-pPb}, for different multiplicity classes in \pPb~collisions at \sqrtSnnE{5.02}. The results are reported as value $\pm$ stat. $\pm$ sys. (uncorr.), where the first error is the statistical uncertainty, the second is the total systematic uncertainty and the value in parentheses indicates the component of uncertainty uncorrelated across multiplicity classes.}
\label{tab:ratios}
\end{table}%
The systematic uncertainty on tracking, track selection, material budget and hadronic interaction cross section are correlated among each particle and its decay products, thus they partially cancel out in the propagation of the error to the final ratio. The residual uncertainties after cancellation are correlated across the event classes. Systematic uncertainties derived from signal extraction and PID selection are uncorrelated. 
\\Based on the results reported in Tab.~\ref{tab:ratios}, one can conclude that no significant multiplicity dependence is observed in the \ks/\pion~and the \ks/p ratios. The 2\ph/(\pip+\pim)~ratio exhibits instead an increasing trend with multiplicity, going from 0.0143 $\pm$ 0.001 in the lowest multiplicity bin to 0.0185 $\pm$ 0.001 in the highest multiplicity class, for a total increase of 29$\%$ with a 2.6$\sigma$ significance. A similar trend with multiplicity is also observed for the 2\ph/(p+\pbar) ratio, which increases by about 24$\%$ with a significance of 1.3$\sigma$ going from 80--100$\%$ to 0--5$\%$. 
\\The increase of the 2\ph/(\pip+\pim)~ratio with multiplicity can be interpreted in the context of strangeness enhancement. The enhancement of \ph-meson (s$\overline{\mathrm{s}}$) production relative to pion has been observed in \PbPb~to follow the enhancement observed for other strange and multi-strange baryons \cite{ref:alice-rsn-PbPb}. In \pPb~the results are in general agreement with the results on $\Xi$/\pion~ and $\Omega$/\pion~ratio \cite{ref:alice-multis-pPb}, that seem to indicate that the strangeness content may control the rate of increase with multiplicity.

Most interesting are the ratios of \ks~and \ph~to charged K, which have been compared to similar measurements in \pp~at \sqrtSE{7} \cite{ref:alice-rsn-pp7TeV} and \PbPb~collisions at \sqrtSnnE{2.76} \cite{ref:alice-rsn-PbPb}, looking for indications of the presence of re-scattering effects in central \pPb~collisions.  
\ks/K and \ph/K in the three collision systems are reported in Fig.~\ref{fig:rsn2ka} as a function of \dndetacube. 
While spanning a smaller range of particle multiplicities, the \ks/K and \ph/K ratios in \pPb~cover within uncertainties the range of values measured in peripheral (40--60$\%$ and 60--80$\%$) \PbPb~and pp collisions. 
\\In order to quantify the evolution of the \pPb~ratios with multiplicity class, the ratios ($y$) have been fitted with a first order polynomial, $y = ax + b$, where $x$ = \dndetacube. Only the statistical and -uncorrelated systematic uncertainties, added in quadrature, have been considered for the purpose of the fit. 
In \pPb~collisions the \ph/K ratio follows the trend from minimum bias pp to peripheral \PbPb~collisions. The linear fit to the \pPb~data returns a positive but small slope parameter, a$_{\ph}$ = 0.008 $\pm$ 0.004. 
A similar fit to the \ks/K ratio in \pPb~instead, results into a negative slope, a$_{\ks}$ = -0.030 $\pm$ 0.018. The pp value for the \ks/K ratio is consistent with the ratio in the lowest multiplicity \pPb~events. 
The slope obtained fitting the \PbPb~data, a'$_{\ks}$  = -0.016 $\pm$ 0.006, is interpreted as due to re-scattering effects in central collisions \cite{ref:alice-rsn-PbPb}. 
The slopes in \PbPb~and \pPb~are compatible within the uncertainties (about 60$\%$ in \pPb~and 27$\%$ for \PbPb), and the decreasing trend in \ks/K
may be a hint of the presence of re-scattering effects in high-multiplicity \pPb~events and indicative for a finite lifetime of the hadronic
phase in \pPb~collisions. Further comparisons with models of \pPb~collisions which include resonances and re-scattering effects would be useful
to distinguish between the different scenarios.

\section{Conclusions}\label{sec:conclusions}
The production of \ks~resonances and \ph~mesons in \pPb~collisions at \sqrtSnnE{5.02} has been measured with the ALICE detector, including multiplicity-dependent transverse momentum spectra, mean transverse momentum and particle ratios to long-lived light-flavoured hadron production. 
The system size dependence of these observables has been studied by comparing the \pPb~results with previous measurements in \PbPb~and \pp~collisions. 
In all collision systems, the mean transverse momentum increases with multiplicity for all particle species. The mass ordering observed in central \PbPb~collisions, where particle with similar mass have similar \meanpt, can be attributed to the presence of radial flow. In \pPb~as well as in \pp~collisions \meanpt~mass ordering is not observed.
The measurement of \meanpt~for other hadronic species could shed more light on whether the observed effect is due to the mesonic (baryonic) nature of the particles, or instead, this behaviour is common to resonances rather than long-lived hadrons. 
Ratios of \ks~and \ph~production to charged K are found to be in agreement with the ratios measured at similar multiplicities in pp and \PbPb~
collisions. The measurements in \pPb~follow the trend observed in \PbPb~within the accessible multiplicity range and the uncertainties.
The \ks/K ratio exhibits a finite negative slope from the lowest to the highest multiplicity \pPb~events, suggestive of a finite lifetime of the
hadronic phase in the small \pPb~system.
%

\newenvironment{acknowledgement}{\relax}{\relax}
\begin{acknowledgement}
\section*{Acknowledgements}

The ALICE Collaboration would like to thank all its engineers and technicians for their invaluable contributions to the construction of the experiment and the CERN accelerator teams for the outstanding performance of the LHC complex.
The ALICE Collaboration gratefully acknowledges the resources and support provided by all Grid centres and the Worldwide LHC Computing Grid (WLCG) collaboration.
The ALICE Collaboration acknowledges the following funding agencies for their support in building and
running the ALICE detector:
State Committee of Science,  World Federation of Scientists (WFS)
and Swiss Fonds Kidagan, Armenia;
Conselho Nacional de Desenvolvimento Cient\'{\i}fico e Tecnol\'{o}gico (CNPq), Financiadora de Estudos e Projetos (FINEP),
Funda\c{c}\~{a}o de Amparo \`{a} Pesquisa do Estado de S\~{a}o Paulo (FAPESP);
National Natural Science Foundation of China (NSFC), the Chinese Ministry of Education (CMOE)
and the Ministry of Science and Technology of China (MSTC);
Ministry of Education and Youth of the Czech Republic;
Danish Natural Science Research Council, the Carlsberg Foundation and the Danish National Research Foundation;
The European Research Council under the European Community's Seventh Framework Programme;
Helsinki Institute of Physics and the Academy of Finland;
French CNRS-IN2P3, the `Region Pays de Loire', `Region Alsace', `Region Auvergne' and CEA, France;
German Bundesministerium fur Bildung, Wissenschaft, Forschung und Technologie (BMBF) and the Helmholtz Association;
General Secretariat for Research and Technology, Ministry of Development, Greece;
National Research, Development and Innovation Office (NKFIH), Hungary;
Department of Atomic Energy and Department of Science and Technology of the Government of India;
Istituto Nazionale di Fisica Nucleare (INFN) and Centro Fermi -
Museo Storico della Fisica e Centro Studi e Ricerche ``Enrico Fermi'', Italy;
Japan Society for the Promotion of Science (JSPS) KAKENHI and MEXT, Japan;
Joint Institute for Nuclear Research, Dubna;
National Research Foundation of Korea (NRF);
Consejo Nacional de Cienca y Tecnologia (CONACYT), Direccion General de Asuntos del Personal Academico(DGAPA), M\'{e}xico, Amerique Latine Formation academique - 
European Commission~(ALFA-EC) and the EPLANET Program~(European Particle Physics Latin American Network);
Stichting voor Fundamenteel Onderzoek der Materie (FOM) and the Nederlandse Organisatie voor Wetenschappelijk Onderzoek (NWO), Netherlands;
Research Council of Norway (NFR);
National Science Centre, Poland;
Ministry of National Education/Institute for Atomic Physics and National Council of Scientific Research in Higher Education~(CNCSI-UEFISCDI), Romania;
Ministry of Education and Science of Russian Federation, Russian
Academy of Sciences, Russian Federal Agency of Atomic Energy,
Russian Federal Agency for Science and Innovations and The Russian
Foundation for Basic Research;
Ministry of Education of Slovakia;
Department of Science and Technology, South Africa;
Centro de Investigaciones Energeticas, Medioambientales y Tecnologicas (CIEMAT), E-Infrastructure shared between Europe and Latin America (EELA), 
Ministerio de Econom\'{i}a y Competitividad (MINECO) of Spain, Xunta de Galicia (Conseller\'{\i}a de Educaci\'{o}n),
Centro de Aplicaciones Tecnológicas y Desarrollo Nuclear (CEA\-DEN), Cubaenerg\'{\i}a, Cuba, and IAEA (International Atomic Energy Agency);
Swedish Research Council (VR) and Knut $\&$ Alice Wallenberg
Foundation (KAW);
Ukraine Ministry of Education and Science;
United Kingdom Science and Technology Facilities Council (STFC);
The United States Department of Energy, the United States National
Science Foundation, the State of Texas, and the State of Ohio;
Ministry of Science, Education and Sports of Croatia and  Unity through Knowledge Fund, Croatia;
Council of Scientific and Industrial Research (CSIR), New Delhi, India;
Pontificia Universidad Cat\'{o}lica del Per\'{u}.
\end{acknowledgement}
                                                                                                      
\bibliographystyle{utphys}   
\bibliography{biblio}


\newpage
\appendix
\section{The ALICE Collaboration}
\label{app:collab}

\bigskip 

J.~Adam$^{\rm 40}$, 
D.~Adamov\'{a}$^{\rm 84}$, 
M.M.~Aggarwal$^{\rm 88}$, 
G.~Aglieri Rinella$^{\rm 36}$, 
M.~Agnello$^{\rm 110}$, 
N.~Agrawal$^{\rm 48}$, 
Z.~Ahammed$^{\rm 132}$, 
S.~Ahmad$^{\rm 19}$, 
S.U.~Ahn$^{\rm 68}$, 
S.~Aiola$^{\rm 136}$
% ...

\bigskip

\bigskip 

\textbf{\Large Affiliation Notes}

\bigskip 

% ...
$^{\rm IV}$ Also at: M.V. Lomonosov Moscow State University, D.V. Skobeltsyn Institute of Nuclear, Physics, Moscow, Russia

\bigskip

\bigskip 

\textbf{\Large Collaboration Institutes}

\bigskip 

% ...
$^{19}$ Department of Physics, Aligarh Muslim University, Aligarh, India\\
% ...
$^{36}$ European Organization for Nuclear Research (CERN), Geneva, Switzerland\\
% ...
$^{40}$ Faculty of Nuclear Sciences and Physical Engineering, Czech Technical University in Prague, Prague, Czech Republic\\
% ...
$^{48}$ Indian Institute of Technology Bombay (IIT), Mumbai, India\\
% ...
$^{68}$ Korea Institute of Science and Technology Information, Daejeon, South Korea\\
% ...
$^{84}$ Nuclear Physics Institute, Academy of Sciences of the Czech Republic, \v{R}e\v{z} u Prahy, Czech Republic\\
% ...
$^{88}$ Physics Department, Panjab University, Chandigarh, India\\
% ...
$^{110}$ Sezione INFN, Turin, Italy\\
% ...
$^{132}$ Variable Energy Cyclotron Centre, Kolkata, India\\
% ...
$^{136}$ Yale University, New Haven, Connecticut, United States\\
% ...

\bigskip 


\end{document}